\documentclass[pra,amsmath,aps,12pt,superscriptaddress,letterpaper,nofootinbib,tightenlines]{revtex4}
\usepackage{graphicx,xspace,units}
\usepackage{float}
\usepackage{amsmath}
\usepackage{amssymb}
\usepackage[normalem]{ulem}
\usepackage{wrapfig}
\usepackage{epsfig}
\usepackage[usenames]{color}
\numberwithin{equation}{section}

\definecolor{MidnightBlue}{cmyk}{0.98,0.13,0,0.43}
\definecolor{DarkGreen}{rgb}{0,0.7,0.1}

\newcommand{\refeq}[1]{{(\ref{eq:#1})}}
\newcommand{\refeqn}[1]{{Eq.~(\ref{eq:#1})}}
\newcommand{\labeleqn}[1]{\label{eq:#1}}
\newcommand{\reffig}[1]{{Fig.~\ref{fig:#1}}}
\newcommand{\be}{\begin{equation}}
\newcommand{\ee}{\end{equation}}

\newcommand{\Lag}{\mathcal{L}}

\newcommand{\curl}{\boldsymbol{\nabla} \times}
\newcommand{\bnabla}{\boldsymbol{\nabla}}
\newcommand{\E}{\mathbf{E}}
\newcommand{\D}{\mathbf{D}}
\newcommand{\B}{\mathbf{B}}
\newcommand{\Hf}{\mathbf{H}}
\newcommand{\A}{\mathbf{A}}
\newcommand{\J}{\mathbf{J}}
\newcommand{\M}{\mathbf{M}}
\newcommand{\N}{\mathbf{N}}

\newcommand{\tV}{\mathbb{V}}
\newcommand{\bI}{\mathcal{I}}
\newcommand{\tI}{\mathbb{I}}
\newcommand{\tA}{\mathbb{A}}
\newcommand{\tB}{\mathbb{B}}
\newcommand{\tC}{\mathbb{C}}
\newcommand{\tD}{\mathbb{D}}
\newcommand{\calE}{\mathcal{E}}

\newcommand{\minmunew}{\left(\frac{1}{\mu(\omega,\vecx)}-1\right)}

\newcommand{\minepnew}{\left(1-\epsilon(\omega,\vecx)\right)}

\newcommand{\dA}{\mathcal{D}\A}

\newcommand{\dJ}{\mathcal{D}\J}
\newcommand{\dJJ}{\left. \dJ\dJ^* \right|_\text{obj}}
\newcommand{\dJJprime}{\left. \dJ'{\dJ'}^{*} \right|_\text{obj}}
\newcommand{\bra}[1]{\langle #1 |}
\newcommand{\ket}[1]{| #1 \rangle}

\newcommand{\tr}{\text{tr }}
\newcommand{\half}{\frac{1}{2}}

\newcommand{\aindex}{\alpha}
\newcommand{\bindex}{\beta}
\newcommand{\cindex}{\gamma}
\newcommand{\abindex}{\aindex \bindex}

\newcommand{\phiout}{\phi^\text{out}}
\newcommand{\phir}{\phi^\text{reg}}

\newcommand{\F}{\mathbb{F}}
\newcommand{\T}{\mathbb{T}}
\newcommand{\f}{\mathcal{F}}
\newcommand{\bS}{\mathcal{S}}

\newcommand{\X}{\mathbb{X}}

\newcommand{\V}{\mathbb{V}}
\newcommand{\Y}{\mathcal{N}}
\newcommand{\Hzero}{\mathbb{H}_0}
\newcommand{\Ham}{H}
\newcommand{\HH}{\mathcal{H}}
\newcommand{\Eh}{\E_0}
\newcommand{\Ein}{\E^\text{in}}
\newcommand{\Eout}{\E^\text{out}}
\newcommand{\Ereg}{\E^\text{reg}}
\newcommand{\Eincc}{\E^{\text{in}*}}
\newcommand{\Eregcc}{\E^{\text{reg}*}}
\newcommand{\EregbQcc}{\Eregcc_{\bindex}}
\newcommand{\EregaP}{\Ereg_{\aindex}}
\newcommand{\EoutaP}{\Eout_{\aindex}}
\newcommand{\EoutkaaP}{\Eout_{\aindex}(\kappa)}
\newcommand{\EoutkabQ}{\Eout_{\bindex}(\kappa)}
\newcommand{\EinaPcc}{\E_{\aindex}^{\text{in}*}}
\newcommand{\EinbQcc}{\E_{\bindex}^{\text{in}*}}

\newcommand{\EinkaaP}{\Ein_{\aindex}(\kappa)}
\newcommand{\EinkabQ}{\Ein_{\bindex}(\kappa)}

\newcommand{\EregbQ}{\Ereg_{\bindex}}
\newcommand{\EoutbQ}{\Eout_{\bindex}}

\newcommand{\Eom}{\E(\omega)}
\newcommand{\Ek}{\E(\kappa)}
\newcommand{\Ehom}{\Eh(\omega)}
\newcommand{\EromaP}{\E_{\aindex}^{\text{reg}}(\omega)}
\newcommand{\ErompaP}{\E_{\aindex}^{\text{reg}}(\omega')}
\newcommand{\EoutomaP}{\Eout_{\aindex}(\omega)}
\newcommand{\EoutombQ}{\Eout_{\bindex}(\omega)}
\newcommand{\ErombQ}{\Er_{\bindex}(\omega)}
\newcommand{\EinombQ}{\Ein_{\bindex}(\omega)}
\newcommand{\ketEk}{\ket{\Ek}}
\newcommand{\ketEom}{\ket{\Eom}}

\newcommand{\ketEromaP}{\ket{\EromaP}}
\newcommand{\ketErompaP}{\ket{\ErompaP}}
\newcommand{\ketEoutomaP}{\ket{\EoutomaP}}
\newcommand{\ketEoutombQ}{\ket{\EoutombQ}}
\newcommand{\ketErombQ}{\ket{\ErombQ}}
\newcommand{\braErombQ}{\bra{\ErombQ}}
\newcommand{\braEinombQ}{\bra{\EinombQ}}
\newcommand{\braEinkaaP}{\bra{\EinkaaP}}
\newcommand{\braErompaP}{\bra{\ErompaP}}

\newcommand{\ketE}{\ket{\E}}
\newcommand{\ketEh}{\ket{\Eh}}

\newcommand{\ketEoutkaaP}{\ket{\EoutkaaP}}

\newcommand{\braEinkabQ}{\bra{\EinkabQ}}

\newcommand{\ketEoutkabQ}{\ket{\EoutkabQ}}

\newcommand{\tG}{\mathbb{G}}
\newcommand{\tGzero}{\tG_0}
\newcommand{\Er}{\E^\text{reg}}

\newcommand{\EraP}{\E_{\aindex}^{\text{reg}}}
\newcommand{\ErkaaP}{\E_{\aindex}^{\text{reg}}(\kappa)}
\newcommand{\EraPcc}{\E_{\aindex}^{\text{reg}*}}

\newcommand{\ErkabQ}{\Er_{\bindex}(\kappa)}

\newcommand{\ketErkaaP}{\ket{\ErkaaP}}

\newcommand{\ketErkabQ}{\ket{\ErkabQ}}

\newcommand{\braErkabQ}{\bra{\ErkabQ}}

\newcommand{\braErkaaP}{\bra{\E_{\aindex}^{\text{reg}}(\kappa)}}

\newcommand{\Mr}{\M^\text{reg}}
\newcommand{\Mrcc}{\M^{\text{reg}*}}
\newcommand{\Mout}{\M^\text{out}}
\newcommand{\Mincc}{\M^{\text{in}*}}
\newcommand{\Nr}{\N^\text{reg}}
\newcommand{\Nrcc}{\N^{\text{reg}*}}
\newcommand{\Nout}{\N^\text{out}}
\newcommand{\Nincc}{\N^{\text{in}*}}

\newcommand{\veck}{\mathbf{k}}
\newcommand{\veckpe}{\veck_\perp}

\newcommand{\vecz}{\mathbf{z}}
\newcommand{\vecy}{\mathbf{y}}
\newcommand{\vecx}{\mathbf{x}}
\newcommand{\vecxpe}{\vecx_\perp}
\newcommand{\hatz}{\hat{\mathbf{z}}}

\newcommand{\vecX}{\mathbf{X}}

\newcommand{\orig}{\mathcal{O}}

\newcommand{\ketxp}{|\vecx'\rangle}
\newcommand{\ketxpp}{|\vecx''\rangle}

\newcommand{\ketxip}{\ket{\vecx_i'}}
\newcommand{\ketxjpp}{\ket{\vecx_j''}}
\newcommand{\ketxonep}{|\vecx_1'\rangle}
\newcommand{\ketxtwop}{|\vecx_2'\rangle}
\newcommand{\ketxonepp}{|\vecx_1''\rangle}
\newcommand{\ketxtwopp}{|\vecx_2''\rangle}
\newcommand{\brax}{\langle \vecx|}
\newcommand{\braxi}{\bra{\vecx_i}}
\newcommand{\braxip}{\bra{\vecx_i'}}
\newcommand{\braxone}{\langle \vecx_1|}
\newcommand{\braxtwo}{\langle \vecx_2|}
\newcommand{\braxonep}{\langle \vecx_1'|}
\newcommand{\braxtwop}{\langle \vecx_2'|}

\newcommand{\Tregreg}{\f^{ee}}

\newcommand{\Tregout}{\f^{ei}}
\newcommand{\Toutreg}{\f^{ie}}
\newcommand{\Toutout}{\f^{ii}}

\newcommand{\thalf}{\tfrac{1}{2}}

\begin{document}

\title{Scattering Theory Approach to Electrodynamic Casimir Forces}

\author{Sahand Jamal Rahi}
\email{sjrahi@mit.edu}
\affiliation{Department of Physics,
Massachusetts Institute of Technology,
77 Massachusetts Avenue, 
Cambridge, MA 02139, USA}
\affiliation{Kavli Institute for Theoretical Physics,
University of California, Santa Barbara, 
Santa Barbara, CA 93106, USA}

\author{Thorsten Emig}
\affiliation{Department of Physics,
Massachusetts Institute of Technology,
77 Massachusetts Avenue, 
Cambridge, MA 02139, USA}
\affiliation{Kavli Institute for Theoretical Physics,
University of California, Santa Barbara, 
Santa Barbara, CA 93106, USA}
\affiliation{Institut f\"ur Theoretische Physik,
Universit\"at zu K\"oln,
Z\"ulpicher Strasse 77,
50937 K\"oln, Germany}
\affiliation{Laboratoire de Physique Th\'eorique et Mod\`eles Statistiques,
CNRS UMR 8626, B\^at.~100,
Universit\'e Paris-Sud,
91405 Orsay cedex,
France}

\author{Noah Graham}
\affiliation{Department of Physics,
Middlebury College,
Middlebury, VT 05753, USA}

\author{Robert L. Jaffe}
\affiliation{Department of Physics, 
Massachusetts Institute of Technology,
77 Massachusetts Avenue, 
Cambridge, MA 02139, USA}
\affiliation{Center for Theoretical Physics,
Laboratory for Nuclear Science,
Massachusetts Institute of Technology,
Cambridge, MA 02139,
USA}

\author{Mehran Kardar}
\affiliation{Department of Physics, 
Massachusetts Institute of Technology,
77 Massachusetts Avenue, 
Cambridge, MA 02139,
USA}
\affiliation{Kavli Institute for Theoretical Physics,
University of California, Santa Barbara, 
Santa Barbara, CA 93106, USA}

\begin{abstract}   

We give a comprehensive presentation of methods for calculating the 
Casimir force to arbitrary accuracy, for any number of objects, 
arbitrary shapes, susceptibility functions, and separations.
The technique is applicable to objects immersed in media other than
vacuum, nonzero temperatures, and spatial arrangements in
which one object is enclosed in another.  Our method combines each
object's classical electromagnetic scattering amplitude
with universal translation matrices,
which convert between the bases used to calculate scattering for each
object, but are otherwise independent of the details of the individual
objects.  The method is illustrated by re-deriving the Lifshitz formula for
infinite half spaces, by demonstrating the
Casimir-Polder to van der Waals cross-over, and by computing the
Casimir interaction energy of two infinite, parallel, perfect
metal cylinders either inside or outside one another.  Furthermore, it
is used to obtain new results, namely the Casimir energies of a sphere
or a cylinder opposite a plate, all with finite permittivity and
permeability, to leading order at large separation.
\end{abstract}
\maketitle


\section{Introduction}

Materials that couple to the electromagnetic field alter the spectrum
of its quantum and thermal fluctuations.  The
resulting change in energy depends on the relative positions of the
objects, leading to a fluctuation-induced force, usually called the
Casimir force.  This force has been the subject of precision
experimental measurements
\cite{Lamoreaux97,Mohideen98,Roy99,Ederth00,Chan01,Chen02,DeKiewiet03,Harber05,
Chen06,Krause07,Decca07,Chen07,Munday07,Chan08,Kim08,Palasantzas08,Munday09}
and can influence the operation of nanoscale devices \cite{Chan01,Capasso07}.

Casimir and Polder calculated the fluctuation-induced force on
a polarizable atom in front of a perfectly conducting plate and
between two polarizable atoms, both to leading order at large
separation, and obtained a simple result depending only on the atoms'
static polarizabilities \cite{Casimir48-1}.  Casimir then extended this
result to his famous calculation of the pressure on two perfectly
conducting parallel plates \cite{Casimir48-2}. Feinberg and Sucher
\cite{Feinberg68,Feinberg70}
generalized the result of Casimir and Polder to include both electric
and magnetic polarizabilities.  Lifshitz, Dzyaloshinskii, and
Pitaevskii extended Casimir's result for parallel plates by
incorporating nonzero temperature, permittivity, and permeability into
a general formula for the pressure on two infinite half-spaces
separated by a gap
\cite{Lifshitz55,Lifshitz56,Lifshitz57,Dzyaloshinskii61,Lifshitz80}.

In order to study Casimir forces in more general geometries, it turns
out to be advantageous to describe the influence of an arrangement of
objects on the electromagnetic field by the way they scatter
electromagnetic waves. In this article we derive and apply a
representation of the Casimir energy, developed in
Refs.~\cite{Emig07,Emig08}, that characterizes each object by its
on-shell electromagnetic scattering amplitude.  The separations and
orientations of the objects are encoded in universal translation
matrices, which describe how a solution to the source-free Maxwell's
equations in the basis appropriate to one object looks when expanded
in the basis appropriate to another.  The translation matrices depend
on the displacement and orientation of coordinate systems, but not on
the nature of the objects themselves.  The scattering amplitudes and
translation matrices are then combined in a simple algorithm that
allows efficient numerical and, in some cases, analytical
calculations of Casimir forces and torques for a wide variety
of geometries, materials, and external conditions.
We will generalize the formalism summarized in \cite{Emig07}
further to show how it applies in a wide variety of
circumstances, including: 
\begin{itemize}
\item
$n$ arbitrarily shaped objects, whose
surfaces may be smooth or rough or may include edges and cusps;
\item
objects with arbitrary linear electromagnetic response, including
frequency-dependent, lossy electric permittivity and magnetic
permeability tensors;
\item
objects separated by vacuum or by a medium with
uniform, frequency-dependent isotropic permittivity and permeability;
\item
zero or nonzero temperature;
\item
and objects outside of one another or
enclosed in each other.
\end{itemize}

These ideas build on a range of previous related work, an inevitably incomplete subset of which is briefly reviewed here:
Scattering theory methods were first applied to the parallel
plate geometry, when Kats reformulated Lifshitz theory in terms of
reflection coefficients \cite{Kats77}. Jaekel and Reynaud derived the
Lifshitz formula using reflection coefficients for lossless infinite
plates \cite{Jaekel91} and Genet, Lambrecht, and Reynaud
extended this analysis to the lossy case \cite{Genet03}. Lambrecht, Maia
Neto, and Reynaud generalized these results to include non-specular
reflection  \cite{Lambrecht06}.

Around the same time as Kats's work, Balian and Duplantier developed a
multiple scattering approach to the Casimir energy for perfect metal
objects and used it to compute the Casimir energy at asymptotically
large separations \cite{Balian77,Balian78} at both zero and nonzero
temperature. In their approach, information about the conductors is
encoded in a local surface scattering kernel, whose relation to more
conventional scattering formalisms is not transparent, and their approach was
not pursued further at the time. One can find multiple scattering
formulas in an even earlier article by Renne \cite{Renne71}, but
scattering is not explicitly mentioned, and the technique is only used
to rederive older results.

Another scattering-based approach has been to express the Casimir
energy as an integral over the density of states of the fluctuating
field, using the Krein formula \cite{Krein53,Krein62,Birman62} to
relate the density of states to the $\bS$-matrix for scattering from
the ensemble of objects.  This $\bS$-matrix is
difficult to compute in general. In studying many-body
scattering, Henseler and Wirzba connected the $\bS$-matrix of a
collection of spheres \cite{Henseler97} or disks \cite{Wirzba99} to
the objects' individual $\bS$-matrices, which are easy to
find. Bulgac, Magierski, and Wirzba combined this result with the
Krein formula to investigate the scalar and fermionic Casimir effect
for disks and spheres \cite{Bulgac01,Bulgac06,Wirzba08}. Casimir
energies of solitons in renormalizable quantum field theories
have been computed using scattering theory techniques that combine
analytic and numerical methods \cite{Graham09}.

Bordag, Robaschik, Scharnhorst, and Wieczorek
\cite{Bordag85,Robaschik87} introduced path integral methods to the
study of Casimir effects and used them to investigate the
electromagnetic Casimir effect for two parallel perfect metal plates.
Li and Kardar used similar methods to study the scalar thermal Casimir
effect for Dirichlet, Neumann, and mixed boundary conditions
\cite{Li91,Li92}. The quantum extension was developed further by
Golestanian and Kardar \cite{Golestanian97,Golestanian98} and was
subsequently applied to the quantum electromagnetic Casimir effect
by Emig, Hanke, Golestanian, and Kardar, who studied the Casimir
interaction between plates with roughness \cite{Emig01} and between
deformed plates \cite{Emig03}.  (Techniques developed to study the
scalar Casimir effect can be applied to the electromagnetic case for
perfect metals with translation symmetry in one spatial direction,
since then the electromagnetic problem decomposes into two scalar
ones.)  Finally, the path integral approach was connected to
scattering theory by Emig and Buescher \cite{Buescher05}.

Closely related to the work we present here is that of Kenneth and
Klich, who expressed the data required to characterize
Casimir fluctuations in terms of the transition $\T$-operator for
scattering of the fluctuating field from the objects \cite{Kenneth06}.
Their abstract representation made it possible to prove
general properties of the sign of the Casimir force.
In Refs. \cite{Emig07,Emig08}, we developed a framework in which this
abstract result can be applied to concrete calculations.  In this
approach, the $\T$-operator is related to the 
scattering amplitude for each object individually, which in turn is
expressed in an appropriate basis of multipoles.  
While the $\T$-operator is in general ``off-shell,'' meaning it has
matrix elements between states with different frequencies, the
scattering amplitudes are the ``on-shell'' matrix elements of this
operator between states of equal frequency.\footnote{Because of this
relationship, these scattering amplitudes are also referred to as
elements of the $T$-\emph{matrix}.  In standard conventions,
however, the $T$-matrix differs from the matrix elements of the
$\T$-operator by a basis-dependent constant, so we will use the term
``scattering amplitude'' to avoid confusion.} In this approach, the
objects can have any shape or material properties, as long as the
scattering amplitude can be computed in a multipole expansion (or
measured).  The approach can be regarded as a
concrete implementation of the proposal emphasized by Schwinger
\cite{Schwinger75} that the fluctuations of the electromagnetic field
can be traced back to charge and current fluctuations on the objects.
This formalism has been applied and extended in a number of Casimir
calculations \cite{Kenneth:2007jk,Milton08-3,Milton08-4,reid-2009,golestanian-2009,Ttira:2009ku}. 

The basis in which the scattering amplitude for each object is
supplied is typically associated with a coordinate system appropriate
to the object.  Of course a plane, a cylinder, or a sphere would be
described in Cartesian, cylindrical, or spherical coordinates, respectively.
However, any compact object can be described, for example, in
spherical coordinates, provided that the matrix of scattering amplitudes
can be either calculated or measured in that coordinate system.
There are a limited number of coordinate systems in which such a
partial wave expansion is possible, namely those for which  the vector
Helmholtz equation is separable. The translation 
matrices for common separable coordinate systems, obtained
from the free Green's function, are supplied in Appendix
\ref{sec:Translation}.  For typical cases, the final computation of the
Casimir energy can be performed on a desktop computer for a wide range
of separations. Asymptotic results at large separation can be obtained
analytically.

The primary limitation of the method is on the distance between
objects, since the basis appropriate to a given object may become
impractical as two objects approach.  For small separations, 
sufficient accuracy can only be obtained if the calculation is taken 
to very high partial wave order.  In the case of two
spheres, the scattering amplitude is available in a spherical basis,
but as the two spheres approach, the Casimir energy is dominated by
waves near the point of closest approach \cite{Schaden:1998zz}.  As
the spheres come into contact an infinite number of spherical waves
are needed to capture the dominant contribution (see Section
\ref{sec:Green} for further discussion).  A particular basis may also
be fundamentally inappropriate at small separations.  For instance, if
the interaction of two elliptic cylinders is expressed in
an ordinary cylindrical basis, when the elliptic cylinders are close
enough one may not fit inside the smallest circular cylinder that
encloses the other.  In that case the 
cylindrical basis would not ``resolve'' the two objects (although an
elliptic cylindrical basis would; see Section \ref{sec:Green}).
Finally, for a variety of conceptual and computational reasons, we are
limited to linear electromagnetic response.

To illustrate this general formulation, we provide some sample
applications, including the closed-form expressions for computing the
interaction of a plate and a sphere with finite, uniform,
frequency-dependent electric permittivity and magnetic permeability.
We present the Casimir interaction energy explicitly at asymptotically
large separations in terms of the zero frequency permittivities and
permeabilities of the objects. Although most experiments have centered
around the sphere-plate configuration
\cite{Lamoreaux97,Mohideen98,Roy99,Chan01,Decca03,Iannuzzi04,Harber05,Chen06,Chen07,Munday07,Krause07},
it is only recently that the force between a dielectric sphere and an
idealized metallic plate has been obtained for all distances
\cite{Emig08-1}. Subsequently, this result has been extended to the
situation where both objects are described by the plasma model
\cite{Canaguier09}. In addition, we present the Casimir interaction
energy of a plate and a cylinder at asymptotically large distances in
terms of the two objects' zero frequency permittivities and
permeabilities.  Results beyond the leading order using our
closed-form formulation are not explicitly included, but all
the essential formulas are contained here. These results extend the
perfect metal cylinder and plate results presented in Ref. \cite{Emig06}.

The article is organized as follows: In Section II we review the
derivation of the ground state energy of a field theory using path
integrals. In Section III we expand the free electromagnetic Green's
functions in terms of regular and outgoing waves, taking into account
that the pairs of waves in the expansion are evaluated with respect to
two different coordinate systems. This analysis yields the translation
matrices.  Section IV provides an overview of elements of
scattering theory we will use, including the connection between the
$\T$-operator and the scattering amplitude.  In Section V the 
path integral expression for the energy
is re-expressed in terms of the results of the preceding two
sections, yielding the main result, \refeqn{Elogdet}. In
Section VI sample applications are presented:  A short derivation of
the Lifshitz formula, the cross-over between van der Waals and Casimir
regimes for two atoms, a general derivation of previous results for
cylinders \cite{Rahi08-2,Dalvit06}, and new results for the energy
between a dielectric sphere or cylinder and a dielectric plane. A discussion
follows in Section VII.

\section{Casimir energy from field theory}
\label{sec:energy-from-field}
\noindent

\subsection{Electromagnetic Lagrangian and action}
\label{sec:LandSem}
 
We consider the Casimir effect for objects without free charges and
currents but with nonzero electric and magnetic susceptibilities. The
macroscopic electromagnetic Lagrangian density is
\be
\Lag = \half(\E\cdot \D-\B\cdot \Hf).
\labeleqn{Lem}
\ee
The electric field $\E(t,\vecx)$ and the magnetic field $\B(t,\vecx)$
are related to the fundamental four-vector potential
$A^\mu$ by $\E= - c^{-1} \partial_t \A - \bnabla A^0$ and
$\B = \curl \A$. We treat stationary objects whose responses to the
electric and magnetic fields are linear. For such materials, the $\D$
and $\B$ fields are related to the $\E$ and $\Hf$ fields by the
convolutions $\D(t,\vecx) = \int_{-\infty}^\infty dt' \,
\epsilon(t',\vecx) \E(t-t',\vecx)$ and $\B(t,\vecx) =
\int_{-\infty}^\infty dt' \, \mu(t',\vecx) \Hf(t-t',\vecx)$ in time.
We consider local, isotropic permittivity and permeability, although
our derivation can be adapted to  apply to non-local and non-isotropic
media simply by substituting the appropriate non-local and tensor
permittivity and permeability functions. A more formal derivation of
our starting point \refeqn{Lem}, which elucidates the causality
properties of the permeability and permittivity response functions, is
given in Appendix \ref{app:Derivation}.

We define the quantum-mechanical energy through the path integral,
which sums all configurations of the electromagnetic
fields constrained by periodic boundary conditions in time 
between $0$ and $T$.  Outside of this time interval the fields are
periodically continued.  Substituting the Fourier expansions of the form
$\E(t,\vecx) = \sum_{n=-\infty}^\infty \E(\omega_n,\vecx) 
e^{-i \omega_n t}$ with $\omega_n = 2\pi n/T$, we obtain the action
\be
S(T) =\half \int_0^T dt\int d\vecx \, \left(\E\cdot \D-\B\cdot \Hf\right)
= \half T \sum_{n=-\infty}^\infty \int d\vecx
\left(
\E^* \cdot  \epsilon \E - \B^* \cdot \mu^{-1} \B
\right),
\labeleqn{Sem1}
\ee
where $\epsilon$, $\E$, $\mu$, and $\B$ on the right-hand side are
functions of position $\vecx$ and frequency $\omega_n$, and we have
used $\D(\omega,\vecx)=\epsilon(\omega,\vecx)\E(\omega,\vecx)$ and
$\Hf(\omega,\vecx) = \tfrac{1}{\mu(\omega,\vecx)} \B(\omega,\vecx)$.

From the definition of the fields $\E$ and $\B$ in terms of the
vector potential $A^\mu$, we have
$\curl \E = i\frac{\omega}{c} \B$, which enables us
to eliminate $\B$ in the action,
\be
S(T) = \half T \sum_{n=-\infty}^\infty \int d\vecx\left[ \E^{*} \cdot
\left(\tI - \frac{c^2}{\omega_n^2}
\curl \curl \right) \E - \frac{c^2}{\omega_n^2} \E^{*} \cdot \tV \, \E\right],
\labeleqn{Sem2}
\ee
where
\be
\tV = \tI \, \frac{\omega_n^2}{c^2}
\left(1-\epsilon(\omega_n,\vecx)\right) + \curl
\left(\frac{1}{\mu(\omega_n,\vecx)}-1\right) \curl
\labeleqn{Vem1}
\ee
is the potential operator and we have restored the explicit frequency
dependence of $\epsilon$ and $\mu$.  The potential operator is nonzero
only at those points in space where the objects are located ($\epsilon
\neq 1$ or $\mu \neq 1$). At small frequencies, typical materials have
$\epsilon>1$ and $\mu\approx 1$, and $\V$ can be regarded as an
attractive potential.

In the functional integral we will sum over configurations of the
field $A^\mu$.  This sum must be restricted by a
choice of gauge, so that it does not include the infinitely
redundant gauge orbits.  We will choose to work in the gauge $A^0=0$,
although of course no physical results depend on this choice.


\subsection{Casimir energy of a  quantum field}
We use standard tools to obtain a functional integral expression for
the ground state energy of a quantum field in a fixed
background described by $\V(\omega,\vecx)$. The overlap between the
initial state $\ket{\E_a}$ of a system with the state $\ket{\E_b}$ after
time $T$ can be expressed as a functional integral with the fields
fixed at the temporal boundaries \cite{Feynman65},
\be
\bra{\E_b} e^{-i \Ham T\hbar} \ket{\E_a} =\int \left.\dA\,
\right|_{^{\E(t=0)=\E_a}_{\E(t=T)=\E_b}}e^{\frac{i}{\hbar}S[T]},
\ee
where $S(T)$ is the action of \refeqn{Sem1} with the time integrals
taken between zero and $T$, and $\Ham$ is the corresponding Hamiltonian.

If the initial and final states are set equal and summed over, the
resulting functional integration defines the
Minkowski space functional integral
\be
\mathcal{Z}(T) \equiv \sum_{a}\bra{\E_a}e^{-i\Ham T/\hbar}\ket{\E_a}
=\tr e^{-i \Ham T/\hbar} = \int \dA \, 
e^{\frac{i}{\hbar} S[T]},
\labeleqn{tracetime}
\ee
which depends on the time $T$ and the background potential
$\tV(\omega,\vecx)$.  The partition function that describes this
system at temperature $1/\beta$ is defined by
\be
Z(\beta) = {\cal Z}(-i\hbar\beta) = \tr e^{-\beta \Ham},
\labeleqn{part}
\ee
and the free energy $F$ of the field is
\be
F(\beta) = -\frac{1}{\beta}\log Z(\beta).
\labeleqn{free}
\ee
The limit $\beta\to\infty$ projects the ground state energy out of the trace,
\be
 \calE_0 =  F(\beta = \infty) = 
-\lim_{\beta\to\infty} \frac{1}{\beta} \log  Z  {(\beta)}.
\labeleqn{E0}
\ee
The unrenormalized energy $\calE_0$ generally depends on the
ultraviolet cutoff, but  cutoff-dependent contributions arise from the
objects individually and do not depend on their separations or orientations.  
Such terms can also arise after renormalization if objects are assumed
to constrain electromagnetic waves with arbitrarily high frequencies
(for example, if the fields are forced to vanish on a surface). Such
boundary conditions should be regarded as artificial idealizations;
in reality, when the wavelengths of the electromagnetic waves become
shorter than the length scales that characterize the interactions of
the material, the influence of the
material on the waves vanishes \cite{Graham03}. Accordingly, the potential 
$\tV$ should vanish for real materials
in the high-frequency limit.   Since we are only interested in energy
\emph{differences}, we can remove these divergences by subtracting the
ground state energy of the system when the objects are in some
reference configuration. In most cases we will take this configuration
to have the objects infinitely far apart, but when calculating Casimir
energies for one object inside another, some other configuration must
be used.  We denote the partition function for this reference
configuration by $\overline Z$. In this way we obtain the Casimir energy,
\be
\calE = -\lim_{\beta\to\infty} \frac{1}{\beta} \log Z(\beta)/\overline
Z(\beta).
\ee
Throughout our calculation of $\calE$, we will thus be able to neglect
any overall factors that are independent of the relative positions and
orientations of the objects.

\subsection{Euclidean Electromagnetic Action}
\label{sec:EucEMaction}
 
By replacing the time $T$ by $-i\hbar\beta$, we transform the
Minkowski space functional integral $\mathcal{Z}(T)$ into the
partition function $Z(\beta)$.  In $A^0=0$ gauge, the result is
simply to replace the frequencies
$\omega_n = \frac{2\pi n}{T}$ in \refeqn{Vem1} by $i\frac{2\pi n}{\hbar
\beta}=ic\kappa_n$, where $\kappa_n$ is
the $n^{\rm th}$  Matsubara frequency divided by $c$. (In other gauges
the temporal component $A^0$ of the vector field must be rotated too.)

The Lagrangian is quadratic, so the modes with different
$\kappa_n$ decouple and the partition function decomposes into a
product of partition functions for each mode.  Since 
the electromagnetic field is real, we have $\E^*(\omega) =
\E(-\omega)$ on the real axis.  We can thus further simplify this
decomposition on the imaginary axis by considering $\kappa\ge 0$ only,
but allowing $\E$ and $\E^*$ to vary independently in the path
integral.  Restricting to positive $\kappa$ is possible because the
response functions $\epsilon(ic\kappa,\vecx)$ and
$\mu(ic\kappa,\vecx)$ are invariant under a change of sign in
$ic\kappa$, as shown Appendix~\ref{app:Derivation}. In the limit
$\beta\to\infty$, the sum $\sum_{n\geq 0}$ turns into an integral 
$\frac{\hbar c \beta}{2\pi}\int_{0}^\infty d\kappa$, and we have
\be
\calE_0 = -\frac{\hbar c}{2\pi} \int_0^\infty d\kappa \,
\log Z(\kappa),
\labeleqn{EKem}
\ee
where
\be
\begin{split}
Z(\kappa) = \int \dA \dA^* \, 
\exp & \left[ -\beta \int d\vecx \,
\E^{*} \cdot \left(\tI+\frac{1}{\kappa^2}
\curl \curl \right) \E +
\frac{1}{\kappa^2} \E^{*}  \cdot \tV(ic\kappa,\vecx) \, \E
 \right],
\end{split}
\labeleqn{ZKem}
\ee
\be
\tV(ic\kappa,\vecx) = \tI \, \kappa^2
\left(\epsilon(ic\kappa,\vecx)-1\right) + \curl
\left(\frac{1}{\mu(ic\kappa,\vecx)} -1 \right) \curl
\,.
\ee
The potential $\tV(ic\kappa,\vecx)$ is real for real $\kappa$, even
though $\epsilon$ and $\mu$ can have imaginary parts for real
frequencies $\omega$.  Our goal is now to manipulate $Z(\kappa)$ in
\refeqn{ZKem} so that it is computable from the scattering properties
of the objects.

\section{Green's function expansions and translation formulas}
\label{sec:Green}

\subsection{The free Green's function}

The free Green's function and its representations in
various coordinate systems are crucial to our 
formalism.  The free electromagnetic field ($\tV=0$) obeys equations
of motion obtained by extremizing the corresponding action, \refeqn{Sem1},
\be
\left(- \tI \, \frac{\omega^2}{c^2} + 
\curl\curl \right)\E(\omega,\vecx) = 0.
\labeleqn{EoMem1}
\ee
We will employ the electromagnetic dyadic Green's function
$\tGzero$, defined by
\be
\left(- \tI \, \frac{\omega^2}{c^2}
+ \curl\curl \right)\tGzero(\omega,\vecx,\vecx') = 
\,\tI \delta^{(3)}\left(\vecx-\vecx'\right),
\labeleqn{gfeq}
\ee
written here
in the position space representation.  It is easy to express $\tGzero$
as a Fourier transform,
\be
\tGzero(\omega,\vecx,\vecx')=\int
\frac{d\veck}{(2\pi)^{3}}\frac{e^{i\veck\cdot(\vecx-\vecx')}}
{k^{2}-(\omega/c+i\epsilon)^{2}}\left(\tI-\frac{c^{2}}{\omega^{2}}\veck
\otimes \veck\right)\,
\labeleqn{G0fourier}
\ee
where the displacement of the singularities at $k=\pm \tfrac{\omega}{c}$
corresponds to outgoing wave boundary conditions at infinity.  By
replacing the factors of $\veck$ by gradients, $\tGzero$ may be
expressed in terms of elementary functions,
\be
\tGzero(\omega,\vecx,\vecx') = 
\left(\tI-\frac{c^{2}}{\omega^{2}}\bnabla \otimes \bnabla'
\right)\frac{e^{i\omega|\vecx-\vecx'|/c}}{4\pi|\vecx-\vecx'|}\,.
\labeleqn{G0easy}
\ee
In this representation it is easy to see that $\tGzero$ is transverse,
{\it i.e.\/}
$\bnabla \cdot \tGzero(\vecx,\vecx',\omega)=
\tGzero(\vecx, \vecx',\omega)
\cdot \overleftarrow \bnabla'
=0$, for $\vecx\ne\vecx'$. $\tGzero$ is
not transverse at $\vecx=\vecx'$, as can be seen by taking the
divergence of \refeqn{gfeq}.

We work in coordinate systems in which we can use separation of
variables and employ a spectral representation of
$\tGzero(\vecx,\vecx',\omega)$.  That is, we  represent the Green's
function through the complete set of non-singular
(``regular''), transverse solutions to the
differential equation, \refeqn{EoMem1},
\be
 \E^{\rm reg}_{\alpha}(\omega,\vecx) = \langle\vecx\ketEromaP,
\ee
represented formally by the eigenstate kets
$\ketEromaP$, 
where the generalized index $\alpha$ labels the scattering channel,
including the polarization.  For example, for spherical wave functions
it represents the angular momentum quantum numbers $(l,m)$ and the
polarization $E$ or $M$.  We will choose to normalize these
states in accord with standard conventions in electromagnetic
scattering theory; as a result they are not necessarily normalized
according to the conventions typically used in quantum mechanics.  A
list of the eigenfunctions for various common bases is given in
Appendix~\ref{sec:Greenexp}.  The Green's functions can be expressed
as the coordinate-space matrix element of the operator
\be
\tGzero(\omega) =
\int_{0}^\infty d\omega' \sum_{\aindex}
\mathcal{C}_{\aindex}(\omega')
\frac{\ketErompaP \, \braErompaP}
{(\omega'/c)^2-(\omega/c+i\epsilon)^2},
\labeleqn{G0intem}
\ee
where the $i \epsilon$ has again been included to implement
outgoing wave boundary conditions, so that the Green's function is
causal.\footnote{The coordinate space matrix element of
\refeqn{G0intem} is transverse for all $\vecx$ and $\vecx'$, and
therefore differs from the Green's function defined in
\refeqn{G0easy} by terms local at $\vecx=\vecx'$.  Since we never
employ $\tGzero$ at coincident points, we ignore this
subtlety \cite{Morse53}.  The use of the retarded Green's
function not only makes sense physically, but is also dictated by the
imaginary-frequency formalism, just as is the case for the response
functions $\epsilon$ and $\mu$. It is the \emph{retarded} response
functions that are analytically continued in the frequency domain to
positive imaginary frequency, as shown in Appendix~\ref{app:Derivation}.}
We use the symbol $\tGzero$ to
represent both the  matrix-valued representation of the Green's
function in position space, \refeqn{gfeq},
and the abstract Hilbert space operator, \refeqn{G0intem}.  The
coefficients  $\mathcal{C}_{\aindex}(\omega')$ are inserted because of our choice of
normalization and ensure that
\be
\int_0^\infty 
d\omega' \sum_\alpha \mathcal{C}_{\aindex}(\omega')
\ketErompaP \, \braErompaP = \tI.
\ee

It is also useful to represent the Green's function in a different
way, in which one of the separable coordinates is identified
as the ``radial'' variable and treated differently from the remaining
coordinates.  We let $\xi_1$ represent this coordinate and denote the
remaining coordinates as $\xi_2$ and $\xi_3$.  We introduce 
the ``outgoing'' solution in $\xi_1$, which is in the same scattering
channel as the corresponding regular solution but obeys outgoing wave
boundary conditions as $\xi_1 \to \infty$.
It is linearly independent of the regular solution.  
The full outgoing solution is then obtained by multiplying
the outgoing solution for $\xi_1$ by the regular solutions for $\xi_2$
and $\xi_3$.  We can then express one of the regular wave functions
in the position space representation of \refeqn{G0intem} as a sum of the
outgoing solution for $\omega$ and the outgoing solution for $-\omega$.
By specifying explicitly which of the two arguments of the Green's
function has a greater value of $\xi_1$, we can carry out the $\omega$
integral for each of these two terms separately by closing the contour
in the appropriate half-plane \cite{Morse53}, and obtain
\be
\tGzero(\omega,\vecx,\vecx') =
\sum_{\aindex} C_{\aindex}(\omega)
\left\{
\begin{array}{l l}
\EoutaP(\omega,\xi_1,\xi_2,\xi_3)\otimes\EraPcc(\omega,\xi'_1,\xi'_2,\xi'_3)
 & \text{if } \xi_1(\vecx) > \xi'_1(\vecx') \\
\EraP(\omega,\xi_1,\xi_2,\xi_3)\otimes\EinaPcc(\omega,\xi'_1,\xi'_2,\xi'_3)
 & \text{if } \xi_1(\vecx) < \xi'_1(\vecx') \\
\end{array}.
\right.
\labeleqn{G0expem}
\ee
In this form, the outgoing wave boundary condition is implemented
explicitly.  Since the Green's function is written as a linear
combination of solutions to the free wave equation, it clearly 
satisfies \refeqn{gfeq} for $\vecx \neq \vecx'$.  The normalization constant
$C_{\aindex}(\omega)$, which is determined using the Wronskian of the
regular and outgoing solutions and the completeness relationship for the
regular solutions in $\xi_2$ and $\xi_3$, sets the correct ``jump
condition'' for $\vecx = \vecx'$.  

The outgoing solution is typically singular at $\xi_1=0$, but 
the Green's function with distinct arguments does not encounter that
region,  because the outgoing function is always evaluated for the
larger argument.  For example, in a spherical system the outgoing
solution could take the form of a spherical Hankel function
$h_l^{(1)}(k r) \sim  \frac{e^{ik r}}{k r}$ with $k=\omega/c$, which
obeys outgoing wave boundary conditions, is singular at the origin,
and is independent of the corresponding regular solution $j_l(k r)$.

We will usually work on the imaginary $k$-axis, in which case we will
encounter the corresponding modified special functions.  We continue
to label these functions as ``regular,''  ``outgoing,'' and
``incoming,'' even though they  now increase exponentially 
for large $\xi_1$ for
incoming and regular waves and decrease exponentially for outgoing
waves.  We also note that it may be convenient to redefine the wave
functions to match the standard form of the corresponding modified
functions, and to assign different phases to the two polarizations.
The prefactor $C_{\aindex}(\omega)$ is then correspondingly redefined
as $C_{\aindex}(\kappa)$ to incorporate these changes.  A list of
Green's function expansions in various common bases
is given in Appendix~\ref{sec:Greenexp}.

For a Cartesian coordinate system some of the previous statements
have to be adapted slightly.  We will take one of the Cartesian
coordinates, say $z$, to be the ``radial'' coordinate, as required
by the context.  For example, $z$ might be the direction normal to
the planar surface of a dielectric.  The solutions are then given
in terms of plane waves, $e^{i k_x x + i k_y y \pm
i\sqrt{(\omega/c)^2-\veckpe^2} z}$, where $\veckpe$ is the momentum
perpendicular to the $\hatz$ direction.  All are regular and all
contribute in the integral representation of \refeqn{G0intem}. After
analytic continuation to imaginary frequency, the free Green's
function in Cartesian coordinates is expressed by the above formula if
we identify outgoing solutions with plane wave functions that are
exponentially decreasing in the $+\hatz$ direction, $e^{i k_x x + i
k_y y - \sqrt{\kappa^2+\veckpe^2} z}$, and regular solutions with
the exponentially growing solutions
$e^{i k_x x + i k_y y + \sqrt{\kappa^2+\veckpe^2} z}$.

The wave functions that appear in the series expansion of the free
Green's functions in \refeqn{G0expem} satisfy
wave equations with frequency $\omega$. The integral representations
in \refeqn{G0intem}, on the other hand,
contain wave functions of all frequencies.  As we will see in 
Sect.~\ref{sec:Scatt}, the ability to express the Casimir energy
entirely in terms of an ``on-shell'' partial wave expansion with fixed
$\omega$ will greatly simplify our calculations.

\subsection{Translation matrices}

We will use the free Green's function described in the previous
subsection to combine the scattering amplitudes for two different
objects.  In this calculation, the one argument of the Green's
function will be located on each object.  As a result, if
\refeqn{G0expem} is written in the basis appropriate to one object, we
will want to ``translate'' one of the scattering solutions to the
basis appropriate to the other object.  The configuration of the two
objects --- either outside of each other, or one inside the other ---
will determine which object has the larger or smaller value of $\xi_1$,
and therefore which solution needs to be expanded in the other basis.

\begin{figure}[ht]
\includegraphics[width=0.95\linewidth]{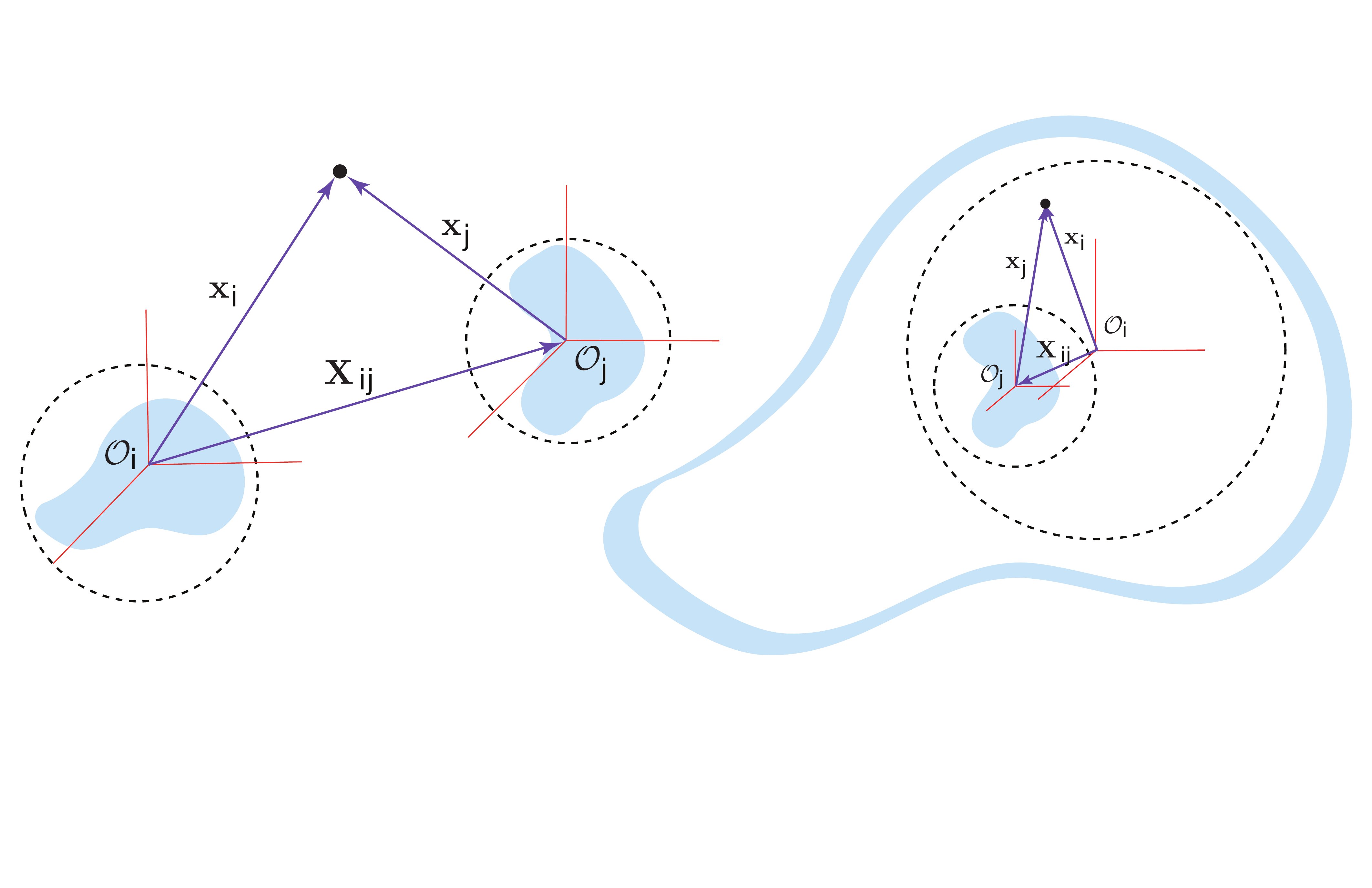}
\caption{(Color online)
Geometry of the outside (left) and inside (right) configurations.
The dotted lines show surfaces separating the objects on which  the
radial variable is constant.  The translation vector
$\vecX_{ij} = \vecx_i - \vecx_j = -\vecX_{ji}$ describes the relative
positions of the two origins.
}
\label{fig:transboth}
\end{figure}

We will make use of two expansions:
\begin{enumerate}
\item
The regular solutions form a complete set no
matter what origin is used to define the decomposition into partial
waves.  Let $\{\EregbQ(\kappa,\vecx_{j})\}$ be the regular solutions
expressed with respect to the origin of coordinates appropriate to
object $j$, ${\cal O}_{j}$.  It must be possible to expand a regular
solution $\EregaP(\kappa,\vecx_{i})$, defined with respect to the
origin ${\cal O}_{i}$ appropriate to object $i$, in terms of the
$\{\EregbQ(\kappa,\vecx_{j})\}$,
\be
\EregaP(\kappa,\vecx_i) = \sum_{\bindex}
\mathcal{V}^{ji}_{\bindex,\aindex}(\kappa,\vecX_{ji})
\EregbQ(\kappa,\vecx_j),
\labeleqn{transregreg}
\ee
where $\vecX_{ij} = -\vecX_{ji} = \vecx_i-\vecx_j$ is
shown in \reffig{transboth}.  Note that $\vecx_{i}$ and $\vecx_{j}$
refer to the same space point $\vecx$, expressed as the displacement 
from different origins.  This expansion will be applicable to the case
of one object inside the other.

\item
The same type of expansion must also exist when the original wave
obeys outgoing boundary conditions \emph{except in a region that
contains the origin ${\cal O}_{i}$}, where $\EoutaP(\kappa,\vecx_{i})$
is singular.  We therefore have the expansion
\be
\labeleqn{transoutreg}
\EoutaP(\kappa,\vecx_i) = \sum_{\bindex}
\mathcal{U}^{ji}_{\bindex,\aindex}(\kappa,\vecX_{ji})
\EregbQ(\kappa,\vecx_j),\,\mbox{for}\,\vecx\notin N({\cal O}_{i})
\ee
where $N({\cal O}_{i})$ is a neighborhood of the origin ${\cal O}_{i}$.
This expansion will be applicable to the case where the objects are
outside each other.
\end{enumerate}

To apply these results to a given geometry, we must be able
to distinguish between regular and outgoing waves over the
whole of each object.  That is, we require there to exist an origin and
a separable coordinate system so that for all points $\vecx$ in one
object and $\vecx'$ in another object, $\xi_1(\vecx)$ is always
greater than $\xi_1(\vecx')$, or vice versa.  Having
$\xi_1(\vecx) > \xi_1(\vecx')$ ensures that the
Green's function is always evaluated by letting $\vecx$ be the
argument of the outgoing wave function and $\vecx'$ be the argument of
the regular wave function.  We therefore require that any two objects
be separated by a surface defined by the locations $\vecx$ where
$\xi_1(\vecx)$ is constant, as shown in \reffig{transboth}.  Depending
on the coordinate system, this surface could be a plane, cylinder,
sphere, etc.

\begin{figure}
\begin{center}
\includegraphics[width=0.65\linewidth]{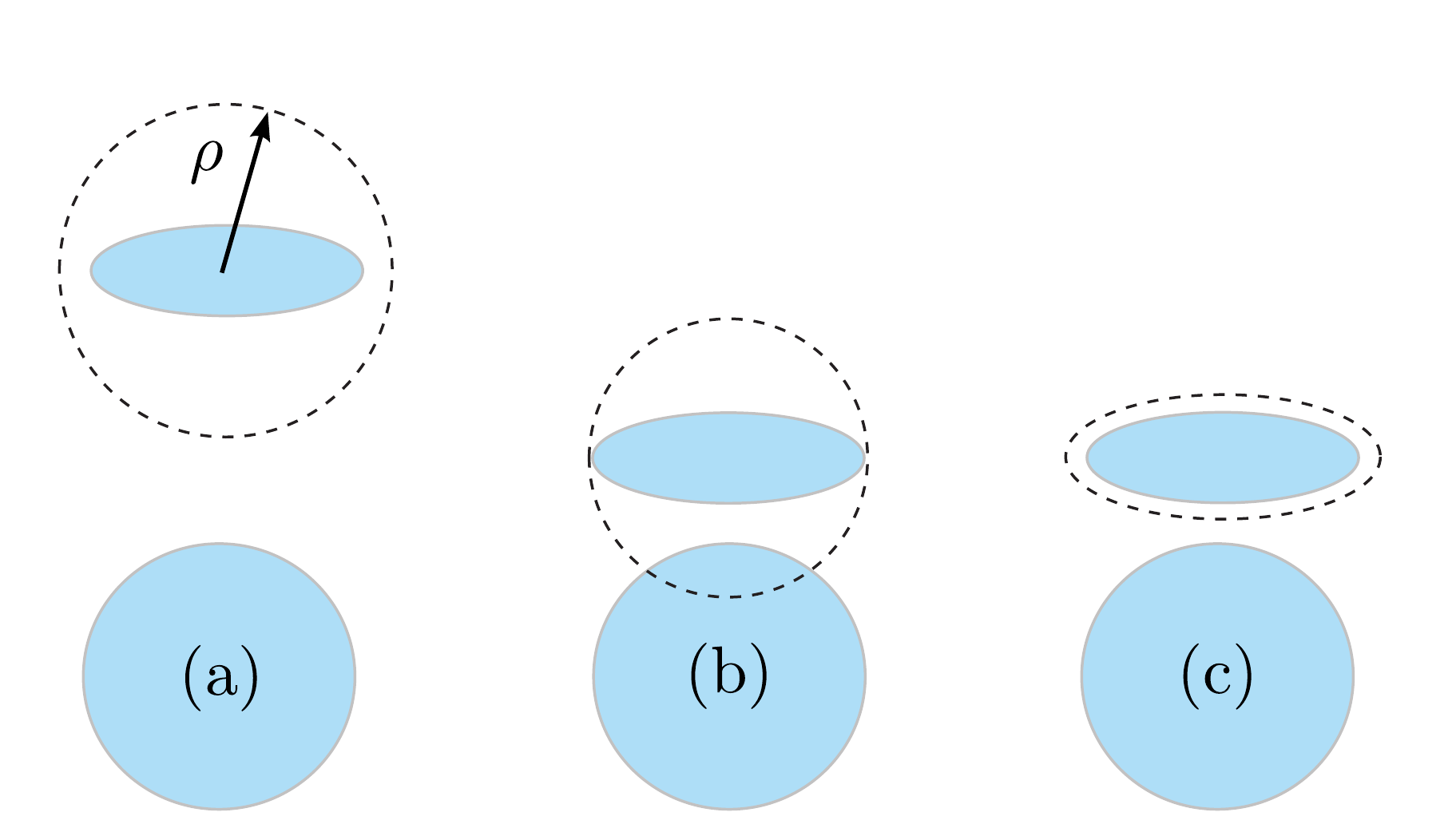}
\caption{An elliptic cylinder approaching another cylinder.  When the
elliptic cylinder is far (a), every point on the cylinder
has smaller radius than any point on the lower cylinder and an expansion using
an ordinary cylindrical basis can be used.  This expansion breaks
down once the elliptic cylinder is close (b), but in that
case an expansion using an elliptic cylindrical basis applies (c).
}
\label{fig:ellipticcylinder}
\end{center}
\end{figure}

The case of an elliptic cylinder and a circular cylinder
illustrates this requirement.  At large distances, the elliptic
cylinder object can be separated from the circular cylinder object by a
circular cylinder of radius $\rho$, as shown in
\reffig{ellipticcylinder}a.  All points on the elliptic cylinder
object have values of $\rho_{1}$ that are smaller than any point on
the circular cylinder object, so in this case we could carry out the
calculation in ordinary cylindrical coordinates.  However, as shown in
\reffig{ellipticcylinder}b, if the separation becomes small enough,
points on the circular cylinder object are closer to the center of the
elliptic cylinder object ({\it i.e.\/} they lie at smaller $\rho_{1}$
than points on the elliptic cylinder object), and our method cannot be
used in ordinary cylindrical coordinates.  However, in \emph{elliptic
  cylindrical} coordinates (see Appendix \ref{ellipticcylinder}), the
surface of the elliptic cylinder object is itself a surface of
constant elliptical radius $\mu_{1}$, so all points on the elliptic
cylinder object have smaller $\mu_{1}$ than any point on the the circular
cylinder object, and our method applies. This case is shown in
\reffig{ellipticcylinder}c.

In a plane wave basis, we would exclude the case of two interlocking
combs \cite{rodriguez-2008}, since each comb has values of $z$ that
are both bigger and smaller than points on the other object, so again
a single assignment of regular and outgoing solutions cannot be made.

When object $j$ lies wholly outside of object $i$,
as shown in the left panel of \reffig{transboth}, in the basis of
object $i$ the point on object $j$ will always have greater $\xi_1$ than the
point on object $i$.  We will therefore need to expand the outgoing
wave in the basis for object $j$.  Since the origin ${\cal
O}_{i}$ is never encountered when the point $\vecx$ lies on
object $j$, the outgoing solutions for $i$ can be expanded in terms of
the regular solutions for object $j$ using \refeqn{transoutreg}. Since
$i$ is also wholly outside $j$, we can also proceed the other way
around and expand the outgoing wave functions in the basis of object
$j$ in terms of regular solutions in the basis of object $i$. This
implies that the translation matrix satisfies
$\mathcal{U}^{ij}=\mathcal{U}^{ji,\dagger}$.
When one object is inside another, as shown in the right panel of
\reffig{transboth}, in the basis of
object $i$, the point on object $j$ will always have smaller $\xi_1$ than the
point on object $i$.  We will therefore need to expand the regular
wave in the basis for object $j$ using \refeqn{transregreg}. In
contrast, we cannot use the expansion of the outgoing wave functions,
because the origin of the inside object may overlap with the origin of
the outside object, in which case the expansion does not converge.

For a Cartesian geometry, the translation matrix is
proportional to $e^{-i\veckpe \cdot \vecX_{ji,\perp} -
\sqrt{\kappa^2+\veckpe^2} X_{ji,z}}$.  It takes this simple form
because plane wave functions are eigenfunctions of the translation
operator.  Then the ``regular'' wave function is evaluated on the
object whose $z$ coordinates are smaller and the outer and inner
objects have larger and smaller $z$ values, respectively.

The criterion for the expansion of the outgoing or regular wave functions
is not topological. Instead, the proximity of the objects and their
origins determines which expansion to use. In practice, it is usually
easy to see which expansion is appropriate for any objects.

After expanding wave functions with respect to another origin using
translation matrices, we can convert the wave functions from one basis
to another, for example from plane wave to spherical or cylindrical wave
functions.  This transformation is useful when the two objects are
best described in different coordinate bases. The needed conversion
matrices are supplied in Appendix~\ref{app:Conversion}. Since it is
more convenient to describe this conversion as a change of basis of
the scattering amplitudes, we will not explicitly consider the
combination of translation and conversion in this derivation, but
instead we will illustrate the change of basis of the scattering
amplitude in the examples.

\subsection{Green's functions and translation matrices}

To obtain the Green's function when one argument, say
$\vecx$, lies on object $i$ and the other argument, say $\vecx'$,
lies on object $j$, we expand $\tGzero(ic\kappa,\vecx,\vecx')$ in
terms of coordinates $\vecx_i$ and $\vecx_j'$ that describe each point
relative to the origin of the body on which it lies.  For the
different situations given above we have
\be
\begin{split}
& \tGzero(ic\kappa,\vecx,\vecx') =  \cr
& \sum_{\aindex,\bindex} C_\beta(\kappa)
\left\{
\begin{array}{l@{~}l}
\EregaP(\kappa,\vecx_i) \otimes \mathcal{U}^{ji}_{\abindex}(\kappa)
\EregbQcc(\kappa,\vecx_j')
&\hbox{if $i$ and $j$ are outside each other} \\
\EregaP(\kappa,\vecx_i) \otimes \mathcal{V}^{ij}_{\abindex}(\kappa)
\EinbQcc(\kappa,\vecx_j')&\left\{\begin{array}{l} 
 \hbox{if $i$ is inside $j$, or} \\ 
 \hbox{if $i$ is below $j$ (plane wave basis)} \end{array}\right.\\
\EoutaP(\kappa,\vecx_i) \otimes \mathcal{W}^{ji}_{\abindex}(\kappa)
\EregbQcc(\kappa,\vecx_j')&\left\{\begin{array}{l} 
 \hbox{if $j$ is inside $i$, or} \\
 \hbox{if $j$ is below $i$ (plane wave basis)}\end{array}\right. \\
\end{array} \right.
\end{split}
\labeleqn{G0cases}
\ee
where $\mathcal{W}^{ji}_{\alpha \beta}(\kappa) =
\mathcal{V}^{ji,\dagger}_{\alpha \beta} (\kappa)
\frac{C_{\aindex}(\kappa)}{C_{\bindex}(\kappa)}$ and $C_\aindex$ is
the normalization constant defined in \refeqn{G0expem}.  We can
express these cases in the consolidated form,
\be
\tGzero(ic\kappa,\vecx,\vecx') =
\sum_{\aindex,\bindex}
C_{\bindex}(\kappa)
\left(\EregaP(\kappa,\vecx_i) ~~ \EoutaP(\kappa,\vecx_i)\right)
\otimes
\left(
\begin{array}{c c}
\mathcal{U}^{ji}_{\abindex}(\kappa) & 
\mathcal{V}^{ij}_{\abindex}(\kappa) \\
\mathcal{W}^{ji}_{\abindex}(\kappa) & 0
\end{array}
\right) \left(
\begin{array}{c}
\EregbQcc(\kappa,\vecx_j') \\
\EinbQcc(\kappa,\vecx_j') 
\end{array}
\right),
\labeleqn{G0shiftlong}
\ee
where only one of the three submatrices is nonzero for any
pair of objects $i$ and $j$ as given in \refeqn{G0cases}.
The expansion can be written more formally as
\be
\tGzero(ic\kappa) = \sum_{\aindex,\bindex}(-C_{\bindex}(\kappa))
\left(\ketErkaaP ~~ \ketEoutkaaP\right)
\X^{ij}_{\abindex}(\kappa)
\left(
\begin{array}{c}
\braErkabQ \\
\braEinkabQ
\end{array}
\right),
\labeleqn{tG0shiftshort}
\ee
where the bras and kets are to be evaluated in position space in the
appropriately restricted domains and the $\X$ matrix is defined, for
convenience, as the negative of the matrix containing the translation
matrices,
\be
\begin{split}
\X^{ij}(\kappa) =
\left(
\begin{array}{c c}
-\mathcal{U}^{ji}(\kappa) & -\mathcal{V}^{ij}(\kappa) \\
-\mathcal{W}^{ji}(\kappa) & 0
\end{array}
\right).
\end{split}
\labeleqn{Xdef}
\ee

The translation matrices for various geometries are provided in
Appendix~\ref{sec:Translation}.

\section{A review of aspects of the classical scattering
of electromagnetic fields}
\label{sec:Scatt}

In this section, we review the key results from scattering theory
needed to compute the scattering amplitude of each body individually.
Comprehensive derivations can be found in
Refs. \cite{Merzbacher98,Newton02}.  The approach we will use was
first developed by  Waterman \cite{Waterman65,Waterman71}, albeit not
in the operator form that is used here.  In the subsequent section we
will then combine these results with the translation matrices of the
previous section to compute $Z(\kappa)$.

\subsection{Electromagnetic scattering}

The Fourier-transformed electromagnetic action of \refeqn{Sem1}
yields the frequency-dependent Maxwell equations:
\be
\curl \E(\omega,\vecx) = i \frac{\omega}{c} \B(\omega,\vecx)\,, \qquad
\curl \frac{1}{\mu} \B(\omega,\vecx) = - i \frac{\omega}{c} \epsilon
\E(\omega,\vecx).
\ee
Combining these two equations, we obtain
\be
(\Hzero + \V(\omega,\vecx))\E(\omega,\vecx) = \frac{\omega^2}{c^2} 
\E(\omega,\vecx),
\labeleqn{EoMem2}
\ee
where
\be
\begin{gathered}
\Hzero = \curl \curl , \\
\V(\omega,\vecx) = \tI \frac{\omega^2}{c^2} \minepnew + \curl \minmunew \curl,
\end{gathered}
\ee
which is the same potential operator as the one obtained by rearranging the
Lagrangian (see \refeqn{Vem1}).   Since the electromagnetic potential
is a differential operator, care must be taken with operator ordering.

The Lippmann-Schwinger equation \cite{Lippmann50}
\be
\ketE = \ketEh - \tGzero \V \ketE
\labeleqn{LS}
\ee
expresses the general solution to \refeqn{EoMem2}.  Here
$\tGzero$ is the free electromagnetic tensor Green's function
discussed in Sec. \ref{sec:Green} and the homogeneous solution $\ketEh$
obeys $\left(-\frac{\omega^2}{c^2}\tI + \Hzero\right) \ketEh = 0$,
which can be chosen to be either a regular or outgoing wave for a
particular frequency $\omega$.  We can iteratively substitute for
$\ketE$ in \refeqn{LS} to obtain the
formal expansion
\be
\begin{split}
\ketE  & = \ketEh - \tGzero \V \ketEh + \tGzero \V \tGzero \V \ketE -
\ldots \\
& = \ketEh - \tGzero \T \ketEh ,
\end{split}
\labeleqn{LSEM}
\ee
where the electromagnetic $\T$-operator is defined as
\be
\T = \V \frac{\tI}{\tI + \tGzero \V} = \V \tG \tGzero^{-1},
\labeleqn{Tem}
\ee
and $\tG$ is the Green's function of the full Hamiltonian,
$\left(-\frac{\omega^2}{c^2} \tI + \Hzero + \V
\right)\tG=\tI$.  We note that  $\T$, $\tGzero$, and $\tG$ are all
functions of frequency $\omega$ and non-local in
space.  As can be seen from expanding $\T$ in \refeqn{Tem} in a power
series, $\T(\omega,\vecx,\vecx')=\bra{\vecx} \T(\omega) \ket{\vecx'}$ is
zero whenever $\vecx$ or $\vecx'$ are not located on an object, {\it
i.e.\/}, where $\V(\omega,\vecx)$ is zero.  This result does not, however,
apply to
\be
\T^{-1} = \tGzero + \V^{-1},
\labeleqn{tinverse}  
\ee
because the free Green's function is nonlocal.

Next we connect the matrix elements of the $\T$-operator between
states with equal $\omega$ to the scattering amplitude $\f$.  In our
formalism, only this restricted subset of $\T$-operator matrix
elements is needed in the computation of the Casimir energy.

\begin{figure}[htb]
\hfill
\includegraphics[width=0.4\linewidth]{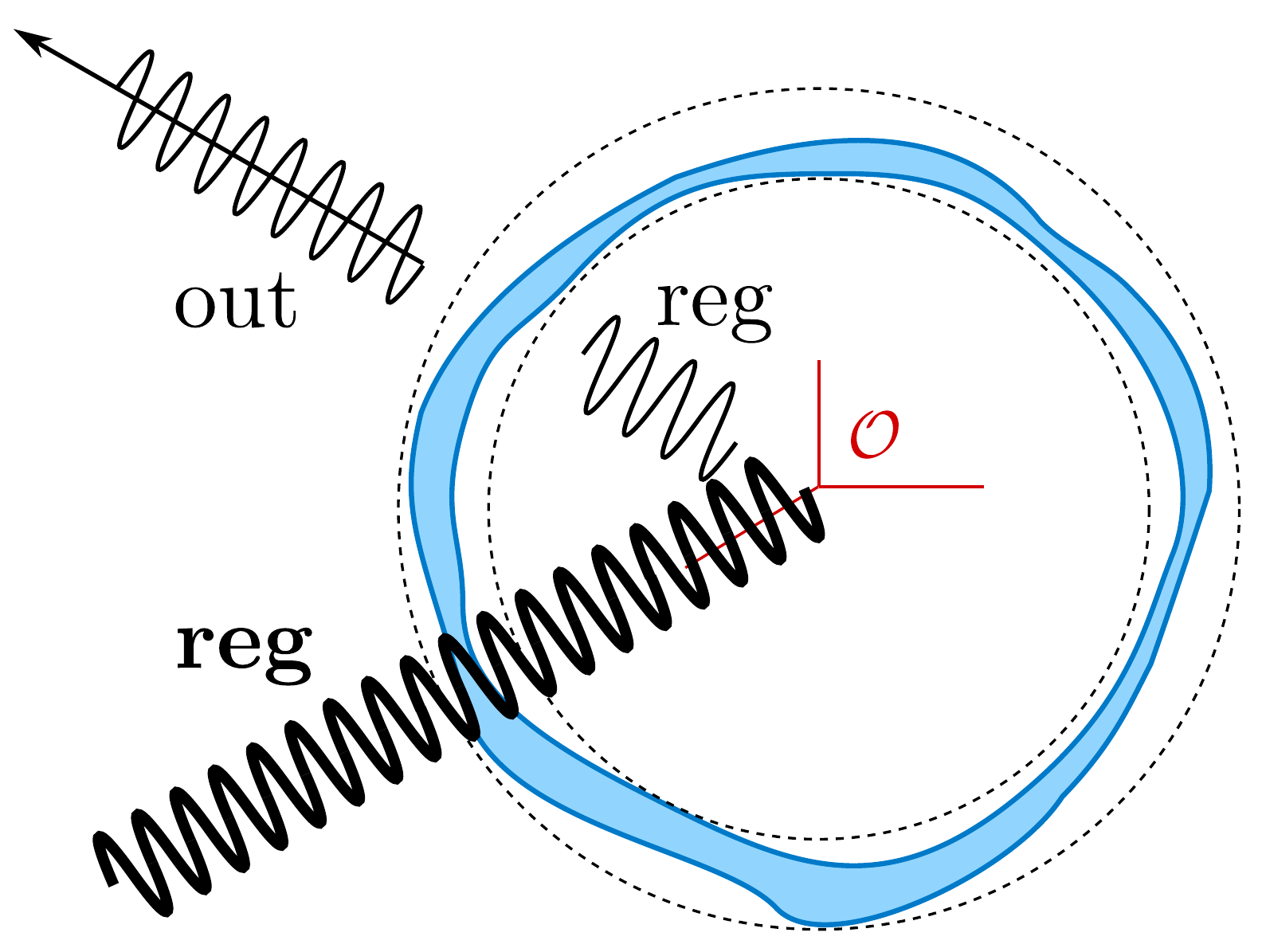}
\hfill
\includegraphics[width=0.4\linewidth]{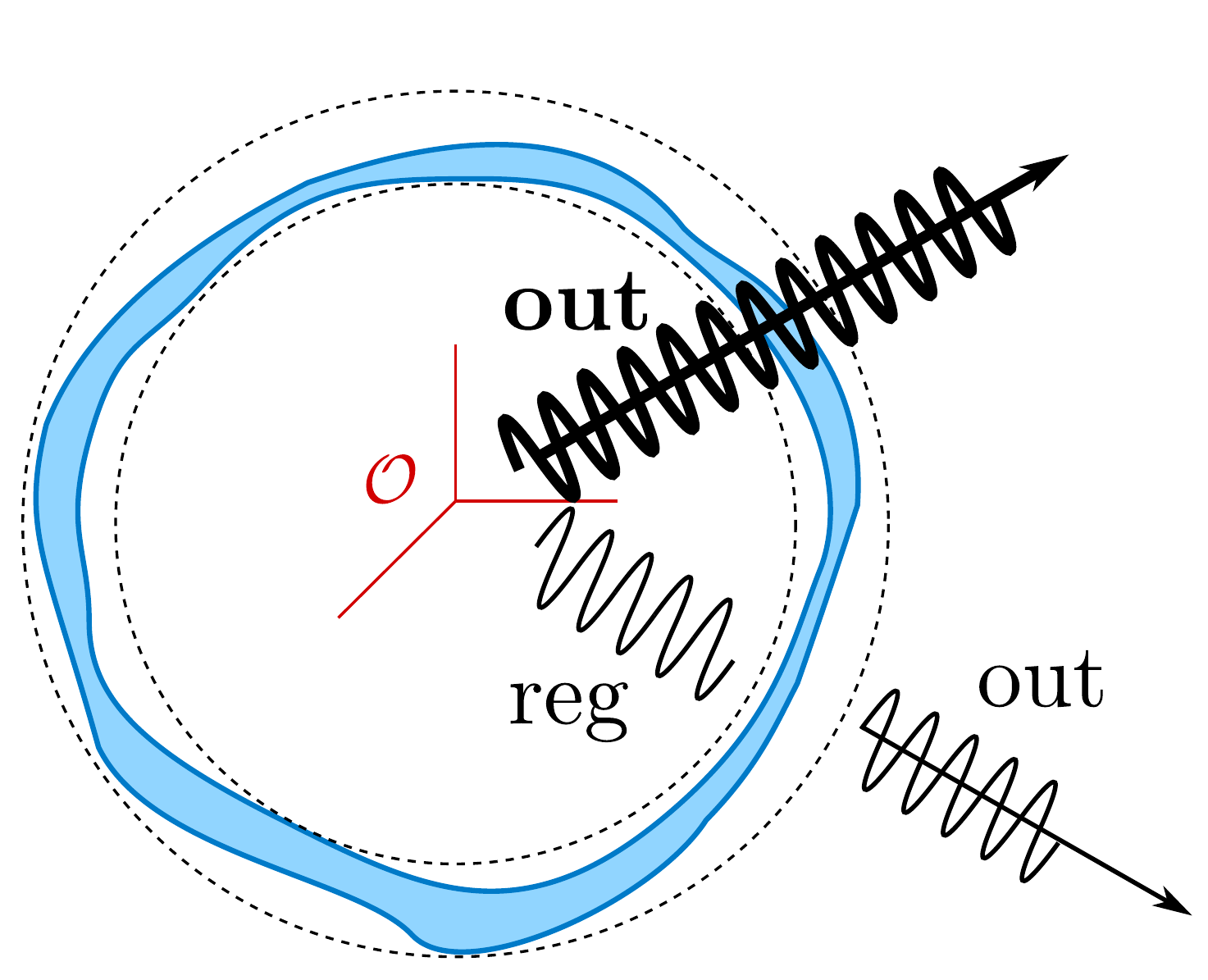}
\hfill
\caption{(Color online) 
The scattering waves for outside scattering (left panel) and inside
scattering (right panel).  In both cases
the homogeneous solution $\Ehom$ is shown in bold.  For outside
scattering, the homogeneous solution is a regular wave,
which produces a regular wave inside the object and an outgoing wave
outside the object.  For inside scattering, the
homogeneous solution is an outgoing wave, which produces a regular
wave inside the object and an outgoing wave outside the object.
}
\label{fig:scatt}
\end{figure}

\subsubsection{Outside scattering}

We consider a scattering process in which a regular wave
interacts with an object and scatters outward, as
depicted in the left panel of \reffig{scatt}.\footnote{
Alternatively, we can set up asymptotically incoming and outgoing
waves on the outside and regular waves inside. The amplitudes of the
outgoing waves are then given by the $\bS$-matrix, which is related to
the scattering amplitude $\f$ by $\f=(\bS-\bI)/2$.  Although these two
matrices carry equivalent information, the scattering amplitude will
be more convenient for our calculation.}
For outside scattering the homogeneous solution $\ketEh$ in
\refeqn{LSEM} is taken to be the regular wave function
$\ketEromaP$. We choose a convenient ``scattering origin'' in the
inside region, consistent with any symmetries of the
problem if possible.

To find the field $\E$ at a coordinate $\vecx$ far enough outside
the object, we use \refeqn{LSEM} in position space and the
expansion in \refeqn{G0expem} for $\tG_0$:
\be
\E(\omega, \vecx)  =  \EregaP(\omega, \vecx) - 
\sum_{\bindex} \EoutbQ(\omega, \vecx) \int
C_{\bindex}(\omega) \EregbQcc(\omega, \vecx') \cdot
\T(\omega, \vecx',\vecx'') \EregaP(\omega, \vecx'') d\vecx'd\vecx''.
\ee
Here ``far enough outside'' means that $\vecx$ has
larger $\xi_1$ than any point on the object, meaning that 
we are always taking the same choice in \refeqn{G0expem}, as
described in Sect.~\ref{sec:Green}.
The equation can be written in Dirac notation, again with the
condition that the domain of the functional Hilbert space 
is chosen appropriately to the type of solution, 
\be
\ketEom = \ketEromaP + \sum_{\bindex} \ketEoutombQ \times
\underbrace{(-1) C_{\bindex}(\omega) \braErombQ \T (\omega)\ketEromaP}_{
\Tregreg_{\bindex, \aindex}(\omega)},
\labeleqn{Temii}
\ee
which defines $\Tregreg_{\beta,\alpha}$ as the exterior/exterior
scattering amplitude (the one evaluated between two
regular solutions).  We will use analogous notation in the other
cases below. 

At coordinates $\vecx$ ``far enough inside'' a cavity of the object,
meaning that $\vecx$ has smaller $\xi_1$ than any point on the object,
we have the opposite case in \refeqn{G0expem} and 
the field $\E$ is given by
\be
\ketEom = \ketEromaP + \sum_{\bindex} \ketErombQ \times
\underbrace{(-1) C_{\bindex}(\omega) \braEinombQ
\T (\omega)\ketEromaP}_{
\Toutreg_{\bindex, \aindex}(\omega)},
\labeleqn{Temoi}
\ee
where again the free states are only defined over the appropriate
domain in position space, and $\Toutreg$ indicates the
interior/exterior scattering amplitude.

\subsubsection{Inside scattering}

In the study of Casimir problems with one object inside the other, it
is useful to imagine a situation that would be difficult to realize in
actual scattering experiments, in which the wave probing the object
originates inside the object and is scattered as a regular wave inside
the object and as an outgoing wave outside, as depicted in the right
panel of \reffig{scatt}.

The situation is expressed mathematically by letting the homogeneous
solution $\ketEh$ in \refeqn{LSEM} be an outgoing wave
$\ketEoutomaP$. The equation can be expressed in condensed form
as before.  Inside the object we have
\be
\ketEom = \ketEoutomaP + \sum_{\bindex} \ketErombQ \times
\underbrace{(-1) C_{\bindex}(\omega) \braEinombQ 
\T (\omega)\ketEoutomaP}_{
\Toutout_{\bindex, \aindex}(\omega)},
\labeleqn{Temoo}
\ee
and outside the object we have
\be
\ketEom = \ketEoutomaP + \sum_{\bindex} \ketEoutombQ \times
\underbrace{(-1) C_{\bindex}(\omega) \braErombQ \T (\omega)
\ketEoutomaP}_{\Tregout_{\bindex, \aindex}(\omega)}.
\labeleqn{Temio}
\ee
%

\subsubsection{Remarks}

We have obtained the scattering amplitude in the basis of free
solutions with fixed $\omega$.  Since one is normally interested in
the scattering of waves outside the object, the scattering amplitude
usually refers to  $\Tregreg$.  We will use a more
general definition, which encompasses all possible combinations of
inside and outside.  The scattering amplitude is always ``on-shell,''
because the frequencies of the bra and ket wave functions are both
equal to $\omega$.  As a result, it is a special case of the
$\T$-operator, which connects wave functions with different $\omega$.

It is usually not practical to calculate the matrix elements by
finding the abstract $\T$-operator and taking its inner products with
free wave functions. Instead, one typically considers an ansatz for
the solutions appropriate for inside or outside scattering in the
various regions, with unknown scattering amplitudes, and then solves
the wave equation, matching the solutions in different
regions at their boundaries.

We will find it convenient to assemble the scattering amplitudes for
inside and outside into a single matrix,
\be
\begin{split}
\F(\kappa) &=
\left(
\begin{array}{c c}
\Tregreg(\kappa) & \Tregout(\kappa) \\
\Toutreg(\kappa) & \Toutout(\kappa)
\end{array}
\right) \\
& =  (-1)C_{\aindex}(\kappa)
\left(
\begin{array}{c c}
\braErkaaP \T(ic\kappa) \ketErkabQ & \braErkaaP \T(ic\kappa)
\ketEoutkabQ \\ \braEinkaaP \T(ic\kappa) \ketErkabQ & \braEinkaaP
\T(ic\kappa) \ketEoutkabQ
\end{array}
\right).
\end{split}
\labeleqn{Temdef}
\ee
Here we have written this expression in terms of modified wave
functions for $\omega=ic\kappa$,
with the corresponding normalization constant, since that is the case
we will use.  The derivations of the scattering amplitudes carry over
directly to this case, with $\kappa$ replaced by $\omega$; for
example, \refeqn{Temii} becomes
\be
\ketEk = \ketErkaaP + \sum_{\bindex} \ketEoutkaaP \times
\underbrace{(-1) C_{\bindex}(\kappa) \braErkabQ \T(ic\kappa)
\ketErkaaP}_{\Tregreg_{\bindex,\aindex}(\kappa)}.
\ee

\section{Partition function in terms of the scattering amplitude}
\noindent

With the tools of the previous two sections, we are now able to 
re-express the Euclidean electromagnetic partition function of
\refeqn{ZKem} in terms of the scattering theory results derived in
Section \ref{sec:Scatt} for imaginary frequency.
We will exchange the fluctuating field $\A$, which
is subject to the potential $\V(ic\kappa,\vecx)$, for 
a free field $\A'$, together with fluctuating currents $\J$ and
charges $-\frac{i}{\omega}\nabla\cdot\J$ that are confined to
the objects.  The sequence of two changes of
variables that will be performed is often referred to as the
Hubbard-Stratonovich transformation in condensed matter physics. 

We multiply and divide the partition function \refeqn{ZKem} by
\be
\begin{split}
W & = \int \dJJ \exp\left[-\frac{T}{\hbar}
\int d\vecx \, \J^*(\vecx) \cdot \V^{-1}(ic\kappa,\vecx)
\J(\vecx) \right]  = \det \V(ic\kappa,\vecx,\vecx')\, ,
\end{split}
\ee
where $\left.\right|_{\rm obj}$ indicates
that the currents are defined only over the objects, {\it i.e.\/} the
domain where $\V$ is nonzero (and therefore $\V^{-1}$ exists),
and we have represented the local potential as a matrix in position space,
$\V(ic\kappa,\vecx, \vecx') = 
\V(ic\kappa,\vecx) \delta^{(3)}(\vecx - \vecx')$.  Our derivation
generalizes straightforwardly to the case of a nonlocal potential
$\V(ic\kappa,\vecx, \vecx')$,
assuming it is still confined to each object individually.

We then change variables in the integration,
$\J(\vecx) = \J'(\vecx) + \frac{i}{\kappa} 
\V(ic\kappa,\vecx) \E(\vecx)$ and
$\J^{*}(\vecx) = {\J'}^{*}(\vecx) + \frac{i}{\kappa} 
\V(ic\kappa,\vecx) \E^{*}(\vecx)$, to obtain
\be
\begin{split}
Z(\kappa) & = \frac{1}{W} \int \dA \dA^* \dJJprime \,
\exp \left[
-\frac{T}{\hbar} \int d\vecx \, \left(\HH \phantom{\frac{1}{1}}
\right. \right. \\ & \left. \left.
+ \left({\J'}^{*}(\vecx) + \frac{i}{\kappa} 
\V(ic\kappa,\vecx) \E^{*}(\vecx)\right) \cdot
\V^{-1}(ic\kappa,\vecx)
\left(\J'(\vecx) + \frac{i}{\kappa} 
\V(ic\kappa,\vecx) \E(\vecx)\right)
\right)
\right],
\end{split}
\labeleqn{EtoJ}
\ee
where
\be
\HH = \E^{*}(\vecx) \cdot \left(\tI + \frac{1}{\kappa^2} \curl \curl \right)
\E(\vecx) 
+ \frac{1}{\kappa^2} \E^{*}(\vecx) \cdot \V(ic\kappa,\vecx) \E(\vecx).
\ee
Next we use a second change of variables, 
$\E(ic\kappa,\vecx) = \E'(ic\kappa,\vecx) - i \kappa \int
d\vecx' \, \tGzero(ic\kappa,\vecx,\vecx') \J'(\vecx')$
and
$\E^{*}(ic\kappa,\vecx) = {\E'}^{*}(ic\kappa,\vecx) - i \kappa \int
d\vecx' \, \tGzero(ic\kappa,\vecx,\vecx') {\J'}^{*}(\vecx')$,
which simplifies \refeqn{EtoJ} to
\be
\begin{split}
Z(\kappa) & = \frac{Z_0}{W} \int
\dJJprime \cr
& \exp\left[
-\frac{T}{\hbar} \int d\vecx d\vecx' \,
{\J'}^{*}(\vecx) \cdot \left(
\tGzero(ic\kappa,\vecx,\vecx')
 + \V^{-1}(ic\kappa,\vecx, \vecx')\right) \J'(\vecx')
\right],
\end{split}
\labeleqn{Zemfull}
\ee
where
\be
Z_0 = \int \dA' \dA'^*
\exp\left[{-\frac{T}{\hbar} \int d\vecx \,
{\E'}^*(\vecx)\cdot
\left(\tI +  \frac{1}{\kappa^2}\curl\curl \right)\E'(\vecx)} \right]
\ee
is the partition function of the free field, which is independent of
the objects. The new partition function of \refeqn{Zemfull} contains a sum
over current fluctuations in place of the original field fluctuations
in \refeqn{ZKem}. The interaction of current fluctuations at different
points $\vecx$ and $\vecx'$ is described by the free Green's function
$\tGzero(ic\kappa,\vecx,\vecx')$ alone. (If the potential
$\V(ic\kappa,\vecx, \vecx')$ is nonlocal, this statement still holds
for two points $\vecx$ and $\vecx'$ on two different objects.)  This
is the expected interaction term. For example, in the static limit
$\kappa=0$, the free Green's function is just the Coulomb interaction
term $\frac{1}{4\pi|\vecx-\vecx'|}$. The inverse potential penalizes
current fluctuations if the potential is small. In vacuum, the
potential vanishes, so current fluctuations are infinitely costly and
thus are not permitted.  But of course the current fluctuations are
already constrained to the objects.

To put the partition function into a suitable form for practical
computations, we will use the results of the previous sections to 
re-express the microscopic current fluctuations
as macroscopic multipole fluctuations, which then can be
connected to the individual objects' scattering amplitudes. 
This transformation comes about naturally once
the current fluctuations are decomposed according to the objects on
which they occur and the appropriate expansions of the
Green's function are introduced.  We begin this process by
noticing that the operator in the exponent of the integrand in
\refeqn{Zemfull} is the negative of the inverse of the $\T$-operator
(see \refeqn{tinverse}), and hence
\be
Z(\kappa) =
Z_0 \, 
\det \V^{-1}(ic\kappa,\vecx,\vecx') \, 
\det \T(ic\kappa,\vecx,\vecx')
\labeleqn{ZemT}
\ee
which is in agreement with a more direct calculation:  Since
$Z_0 = \det \tG_0(ic\kappa,\vecx,\vecx')$ and
$Z(\kappa) = \det \tG(ic\kappa,\vecx,\vecx')$,
 we only need to take the determinant of \refeqn{Tem} to arrive at the
result of \refeqn{ZemT}.

Both $Z_0$ and $\det \V^{-1}(ic\kappa,\vecx)$ are independent of the
separation of the objects, since the former is simply the free Green's
function, while the latter is diagonal in $\vecx$.  Even a nonlocal
potential $\V(ic\kappa,\vecx,\vecx')$ only connects points within the same
object, so its determinant is also independent of the objects'
separation.  Because these determinants do not depend on
separation, they will be canceled by a reference partition function in
the final result.  We are thus left with the task of computing the
determinant of the $\T$-operator.

\subsection{From the determinant of the T-operator to the scattering amplitude}

As has been discussed in Sec. \ref{sec:Scatt}, the $\T$-operator
$\T(ic\kappa,\vecx,\vecx')$ is not diagonal in the spatial
coordinates. Its determinant needs to be taken over the spatial
indices $\vecx$ and $\vecx'$, which are restricted to the objects
because the fluctuating currents $\J(\vecx)$ in the functional
integrals are zero away from the objects.  This determinant also runs
over the ordinary vector components of the electromagnetic $\T$ operator.

A change of basis to momentum space does not help in computing
the determinant of the $\T$-operator, even though it does
help in finding the determinant of the free Green's function. One
reason is that the momentum basis is not orthogonal over the domain of
the indices $\vecx$ and $\vecx'$, which is restricted to the objects. In
addition, a complete momentum basis includes not only all directions
of the momentum vector, but also all magnitudes of the momenta. So, in
the matrix element $\bra{\E_\veck} \T(\omega) \ket{\E_{\veck'}}$ the
wave numbers $k$ and $k'$ would not have to match, and could also
differ from $\omega/c$.  That is, the matrix elements could be
``off-shell.''  Therefore, the $\T$-operator could not simply be treated
as if it was the scattering amplitude, which is the on-shell representation of
the operator in the subbasis of frequency $\omega$
(see Sec. \ref{sec:Scatt}), and is significantly easier to calculate.
Nonetheless, we will see that it is possible to express the Casimir
energy in terms of the on-shell operator only, by remaining in the
position basis.

From \refeqn{Tem}, we know that the inverse of the $\T$-operator
equals the sum of the free Green's function and the inverse of the
potential.  Since the determinant of the inverse operator is the
reciprocal of the determinant, it is expedient to start with the
inverse $\T$-operator.  We then separate the basis involving all the
objects into blocks for the $n$ objects.  In a schematic notation, we have
\be
[\brax \T^{-1} \ketxp] = \left(
\begin{array}{c|c|c}
[\braxone \T_1^{-1} \ketxonep] & 
[\braxone \tGzero \ketxtwop] & \cdots \\ \hline
[\braxtwo \tGzero \ketxonep] & 
[\braxtwo \T_2^{-1} \ketxtwop] & \cdots \\ \hline
\cdots & \cdots & \cdots
\end{array}
\right),
\ee
where the $ij^{\rm th}$ submatrix refers to $\vecx\in$ object $i$ and
$\vecx'\in$ object $j$ and $\vecx_i$ represents a point in object $i$
measured with respect to some fixed coordinate system. Unlike the position
vectors in Sec. \ref{sec:Green}, at this point the subscript of
$\vecx_i$ does not indicate the origin with respect to which the
vector is measured, but rather the object on which the point
lies. Square brackets are used to remind us that we are considering
the entire matrix or submatrix and not a single matrix element.  We
note that the operators $\T$ and $\tGzero$ are functions of
$ic\kappa$, but for simplicity we suppress this argument throughout
this derivation.  When the two spatial indices lie on different
objects, only the free Green's function remains in the off-diagonal
submatrices.  Even if the potential $\V(ic\kappa,\vecx,\vecx')$ is
nonlocal in space, it does not connect points on different
objects. It follows that the inverse of the potential is block
diagonal in position space, where each block involves points on the
same object, {\it i.e.\/},   $\langle  \vecx_{i}|\V^{-1}|
\vecx'_{j}\rangle=0$ for $i\ne j$.

Next, we multiply $\T^{-1}$ by a reference $\T$-operator $\T_\infty$
without off-diagonal submatrices, which can be interpreted as the
$\T$-operator at infinite separation,
\be
\begin{split}
& [\brax \T_\infty  \T^{-1} \ketxpp] = \\
& \left(
\begin{array}{c|c|c}
[\langle \vecx_1 \ketxonepp] &
[\int d\vecx_1' \, \braxone \T_1 \ketxonep \braxonep \tGzero \ketxtwopp] &
\cdots \\ \hline
[\int d\vecx_2' \, \braxtwo \T_2 \ketxtwop \braxtwop \tGzero \ketxonepp] &
[\langle \vecx_2 \ketxtwopp] &
\cdots \\ \hline
\cdots & \cdots & \cdots
\end{array}
\right).
\end{split}
\labeleqn{TTinv}
\ee
Each off-diagonal submatrix $[\int d\vecx_i' \braxi \T_i \ketxip \braxip
\tGzero \ketxjpp]$ is the product of the $\T$-operator of object $i$,
evaluated at two points $\vecx_i$ and $\vecx_i'$ on that object,
multiplied by the free Green's function, which connects $\vecx_i'$
to some point $\vecx_j''$ on object $j$.

Now we shift all variables to the coordinate systems of the objects on
which they lie.  As a result, the index on a position vector
$\vecx_i$ now refers to the object $i$ on which the point lies 
\emph{and} to the coordinate system with origin $\orig_i$ in which the
vector is represented, in agreement with the notation of 
Sec. \ref{sec:Green}.  The off-diagonal submatrices in
\refeqn{TTinv} can then be rewritten using \refeqn{tG0shiftshort} as,
\be
\sum_{\aindex, \bindex}
\left[
\left(\braxi \T_i\ketErkaaP ~~ \braxi \T_i\ketEoutkaaP \right)
\X^{ij}_{\abindex}
\left(
\begin{array}{c}
\braErkabQ \vecx_j'' \rangle \\
\braEinkabQ \vecx_j'' \rangle
\end{array}
\right) (-C_{\bindex}(\kappa))
\right].
\ee

The matrix $[\brax \T_\infty \T^{-1} \ketxpp]$ has the structure 
$\tI + \tA\tB$, where

\be
\begin{split}
&\tA = \sum_{\aindex} \\
&\left(
\begin{array}{c c | c  c |c c }
0 & 0 & 
\left[\braxone \T_1 \ketErkaaP \X^{12}_{\abindex}\right]
& \left[\braxone \T_1 \ketEoutkaaP \X^{12}_{\abindex}\right]
& \cdots & \\
\hline
\left[\braxtwo \T_2 \ketErkaaP \X^{21}_{\abindex}\right]
& \left[\braxtwo \T_2 \ketEoutkaaP \X^{21}_{\abindex}\right]
& 0 & 0 & \cdots  \\ \hline
\cdots & \cdots & \cdots & \cdots & \cdots
\end{array}
\right)
\end{split}
\ee
and
\be
\tB = \left(
\begin{array}{c | c | c }
\left[-C_{\bindex}(\kappa) \braErkabQ \vecx_1'' \rangle \right] & 0 
& \cdots \\
\left[-C_{\bindex}(\kappa) \braEinkabQ \vecx_1'' \rangle \right] & 0 
& \cdots \\ \hline
0 & \left[-C_{\bindex}(\kappa) \braErkabQ \vecx_2'' \rangle \right] 
& \cdots\\
0 & \left[-C_{\bindex}(\kappa) \braEinkabQ \vecx_2'' \rangle \right] 
& \cdots  \\ \hline
\cdots  & \cdots   & \cdots
\end{array}
\right)
\ee
and the matrix multiplication now encompasses both the object index
and the partial wave index $\bindex$.  Although the same symbols are
used for each wave function, the bases (spherical, planar, etc.)
can be chosen differently for each object.

Using Sylvester's determinant formula
$\det(\tI+\tA\tB)=\det(\tI+\tB\tA)$, we see that 
the determinant is unchanged if we
replace the off-diagonal submatrices in \refeqn{TTinv} by
\be
\left[
\sum_{\bindex}
(-1)C_{\aindex}(\kappa)
\left(
\begin{array}{c c}
\braErkaaP \T_i \ketErkabQ & 
\braErkaaP \T_i \ketEoutkabQ \\
\braEinkaaP \T_i \ketErkabQ & 
\braEinkaaP \T_i \ketEoutkabQ
\end{array}
\right)
\X^{ij}_{\bindex, \cindex}
\right].
\labeleqn{TW}
\ee
With this change, the diagonal submatrices in \refeqn{TTinv} become
diagonal in the partial wave indices rather than in position
space. The matrix elements of the $\T$-operator are the scattering
amplitudes, which can be obtained from ordinary scattering calculations,
as demonstrated in Sec. \ref{sec:Scatt}.   The first matrix in
\refeqn{TW}, including the prefactor $(-1)C_{\aindex}(\kappa)$, is
$\F_i(\kappa)$, the modified scattering amplitude of object $i$, defined in
\refeqn{Temdef}.

Putting together Eqs. \refeq{EKem}, \refeq{ZKem}, \refeq{ZemT}, and
\refeq{TTinv}, we obtain
\be
\mathcal{E} = \frac{\hbar c}{2\pi} \int_0^\infty d\kappa 
\log \det (\mathbb{M} \mathbb{M}_\infty^{-1}),
\labeleqn{Elogdet}
\ee
where
\be
\mathbb{M} =
\left(
\begin{array}{c c c c}
\F_1^{-1} & \X^{12} & \X^{13} & \cdots \\
\X^{21}    & \F_2^{-1} & \X^{23} & \cdots \\
\cdots & \cdots & \cdots & \cdots
\end{array}
\right)
\ee
and $\mathbb{M}^{-1}_\infty$ is a block diagonal matrix
$\text{diag}(\F_1 ~~\F_2 ~\cdots)$.

Using the block determinant identity 
\be
\det
\left(
\begin{array}{c c}
\tA & \tB \\
\tC & \tD
\end{array}
\right)
=
\det \left(\tA\right)
\det \left(\tD-\tC \tA^{-1} \tB\right)
=
\det \left(\tD\right)
\det \left(\tA-\tB \tD^{-1} \tC \right),
\labeleqn{detblock}
\ee
we can simplify this expression for the case of the interaction
between two objects,
\be
\mathcal{E} = \frac{\hbar c}{2\pi} \int_0^\infty d\kappa \log \det
\left(\tI - \F_a\X^{ab}\F_b\X^{ba}\right).
\labeleqn{Elogdet2}
\ee

Usually, not all of the submatrices of $\F$ and $\X$ are actually
needed for a computation. For example, if all objects are outside of
one another, only the submatrices $\Tregreg$ of the scattering amplitude
that describe outside reflection are needed.  If there are only two
objects, one inside another, then only the inside reflection submatrix
$\Toutout$ of the outside object and the outside reflection submatrix
$\Tregreg$ of the inside object are needed.

In order to obtain the free energy at nonzero temperature instead of
the ground state energy, we do not take the limit $\beta\to\infty$ in
\refeqn{E0} \cite{Lifshitz55}.  Instead, the integral $\frac{\hbar
c}{2\pi} \int_0^\infty d\kappa$ is replaced everywhere by $\frac{1}{\beta}
\sum_{n}'$, where $c \kappa_n=\frac{2\pi n}{\hbar\beta}$ with
$n=0,1,2,3\ldots$ is the $n$th Matsubara frequency.  A careful
analysis of the derivation shows that the zero frequency mode is
weighted by $1/2$ compared to the rest of the terms in the sum; this
modification of the sum is denoted by a prime on the summation
symbol. The factor of $1/2$ comes about because the fluctuating
charges or currents have to be real for zero frequency. Thus, for
$\kappa_0$, the expressions on the right hand side of
\refeqn{ZemT} should be placed under a square
root.  (For a complex field, both signs of the integer $n$ would be
included separately, and $n=0$ would be included once, with the normal
weight.)

If the medium between the objects is not vacuum but instead has
permittivity $\epsilon_m(ic\kappa)$ and magnetic
permeability $\mu_m(ic\kappa)$ different from unity, then the free
Green's function is multiplied by $\mu_m(ic\kappa)$, and its argument
$\kappa$ is replaced by $n_m(ic\kappa)\kappa$, where
$n_m(ic\kappa)=\sqrt{\epsilon_m(ic\kappa)\mu_m(ic\kappa)}$ is the
medium's index of refraction.  Effectively, this change just scales
all frequency dependencies in the translation matrices $\X(\kappa)$,
which become $\X\left(n_m(ic\kappa)\kappa\right)$.  Furthermore, the
scattering amplitudes absorb the factor $\mu_m(ic\kappa)$ from the
free Green's function and change non-trivially, {\it i.e.\/}
not just by some overall factor or a scaling of the frequency.  They
have to be computed with the nonzero electric and magnetic
susceptibilities of the medium.

\section{Applications}
\label{sec:Applications}

Our technique for calculating the Casimir energy applies to a wide
range of situations.  In this section we demonstrate the method
through a variety of examples.

\subsection{London and Casimir-Polder
interaction between two atoms}

As a simple example, we re-derive the interaction between two
identical neutral atoms in the ground state \cite{Casimir48-1}.  The
atoms are described in a two-state approximation. Within this
approximation, the electric dipole polarizability of the atoms is
given by
\begin{equation}
  \label{eq:alpha_atom}
  \alpha^E=\frac{e^2}{m} \frac{f_{10}}{\omega_{10}^2-\omega^2} \, ,
\end{equation}
where $e$ is the electron charge, $m$ is the mass, $f_{10}$ is the
oscillator strength of the $0\to 1$ transition, and $\omega_{10}$
is the frequency of that transition. We perform a Wick rotation
$\omega \to i c \kappa$ and set $\kappa=u/d$, where $d$ is the
distance between the atoms. By introducing the characteristic length scale
$d_{10}=c/\omega_{10}$ and the static electric polarizability 
\begin{equation}
  \label{eq:alpha_0}
  \alpha_0 = f_{10} r_0 d_{10}^2 \, ,
\end{equation}
where $r_0=e^2/(mc^2)\approx 10^{-15}$m is the classical electron
radius, the polarizability can be written as
\begin{equation}
  \label{eq:alpha_rescaled}
  \alpha^E(u) = \frac{(d/d_{10})^2\alpha_0}{(d/d_{10})^2+u^2} \, .
\end{equation}

In the isotropic-dipole approximation, the only nonzero element of the
scattering amplitude of the atom is given in terms of the electric
dipole polarizability as
\begin{equation}
  \label{eq:T-atom}
  \Tregreg_{1mE,1mE} = \frac{2}{3} \, \alpha^E  \kappa^3 
\end{equation}
for $m=-1,\,0,\, 1$. The atoms are assumed to have no magnetic
polarizability. Using the general expression of \refeqn{Elogdet2}
for the interaction energy between two objects, we get
\begin{eqnarray}
\label{eq:energy-atoms-general}
\mathcal{E} &=& \frac{\hbar c}{2\pi d} \int_0^\infty \!\! du \log \left[
\left(1-4(1+u)^2 e^{-2u} \frac{\left(\alpha^E\right)^2}{d^6} \right)
\left(1-(1+u+u^2)^2e^{-2u} \frac{\left(\alpha^E\right)^2}{d^6}
\right) \right]\\
&=& - \frac{\hbar c}{\pi d^7} 
\int_0^\infty \!\! du \,\left(\alpha^E(u)\right)^2 \, 
(3+6u+5u^2 +2u^3+u^4) \, e^{-2u} \, , 
  \label{eq:energy-atoms-general-expanded}
\end{eqnarray}
where we have expanded the logarithm assuming $\alpha^E(u) \ll d^3$, so
that the interaction energy is proportional to the product of the
polarizabilities of the atoms. It is instructive to consider two
limits. First, assume that $d\ll d_{10}$. By the change of variable
$u=(d/d_{10})z$, one easily finds that in
\refeqn{energy-atoms-general-expanded}, only the leading constant
term of the polynomial in $u$ has to be considered, and the exponential
factor can be ignored. The integral is convergent at large $u$ due to
the behavior of the polarizability $\alpha^E(u)$ at large $u$. The 
integral over $z$ yields, to leading order in $d/d_{10}$, the energy
\begin{equation}
  \label{eq:E_London}
  \mathcal{E}_{\rm L} = - \frac{3}{4} \, \hbar \omega_{10} \,
  \frac{\alpha_0^2}{d^6} \, ,
\end{equation}
which is the well-known London interaction \cite{London}.

In the opposite limit $d\gg d_{10}$ retardation is important.  From
\refeqn{alpha_rescaled} we see that the frequency ($u$)
dependence of the polarizability now can be neglected, so that $\alpha
\approx \alpha_0$. In this retarded limit, the polynomial and exponential
in $u$ in \refeqn{energy-atoms-general-expanded} are important, and
integration yields the energy
\begin{equation}
  \label{eq:energy_CP}
  \mathcal{E}_{\rm CP} = - \frac{23}{4\pi} \, \hbar c \,
  \frac{\alpha_0^2}{d^7} \, ,
\end{equation}
which is known as the Casimir-Polder interaction \cite{Casimir48-1}. 

\begin{figure}[ht]
\includegraphics[scale=0.6]{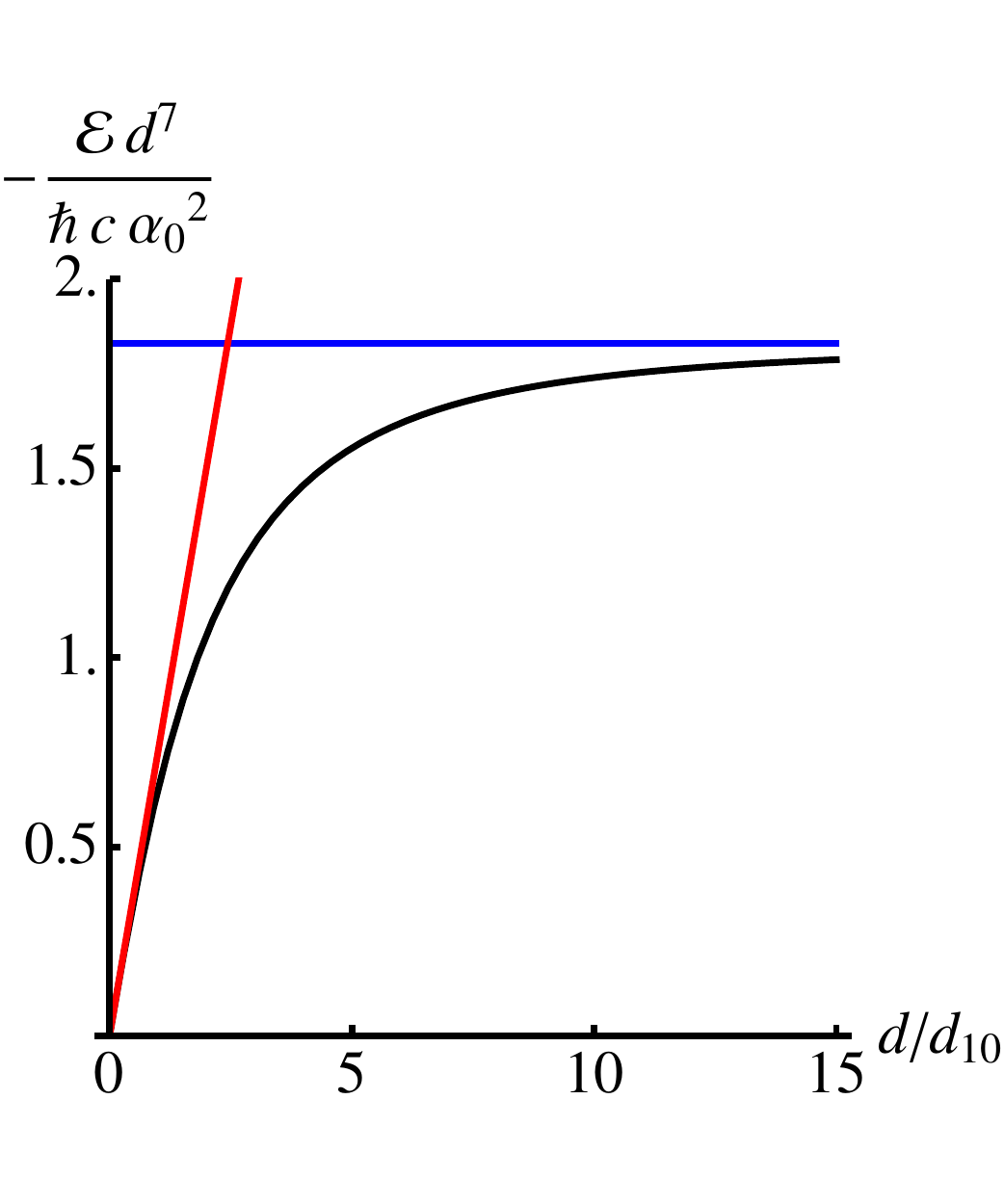}
\caption{(Color online) Interaction energy $\mathcal{E}$ of two
identical atoms, \refeqn{energy-atoms-general-expanded}, as a
function of their separation $d$. The curve shows the crossover
between the London interaction ($d\ll d_{10}=c/\omega_{10}$),
\refeqn{E_London}, and the Casimir-Polder 
interaction ($d\gg d_{10}$), \refeqn{energy_CP}.}
\label{fig:CP-London}
\end{figure}

For general distances $d$, the interaction between the atoms can be
computed numerically; to quadratic order in the polarizability it is
given by the integral of
\refeqn{energy-atoms-general-expanded}. The numerical result
and the two limiting forms of the interaction are shown in
\reffig{CP-London}.

\subsection{Derivation of the Lifshitz formula}

\begin{figure}[ht]
\includegraphics[scale=0.45]{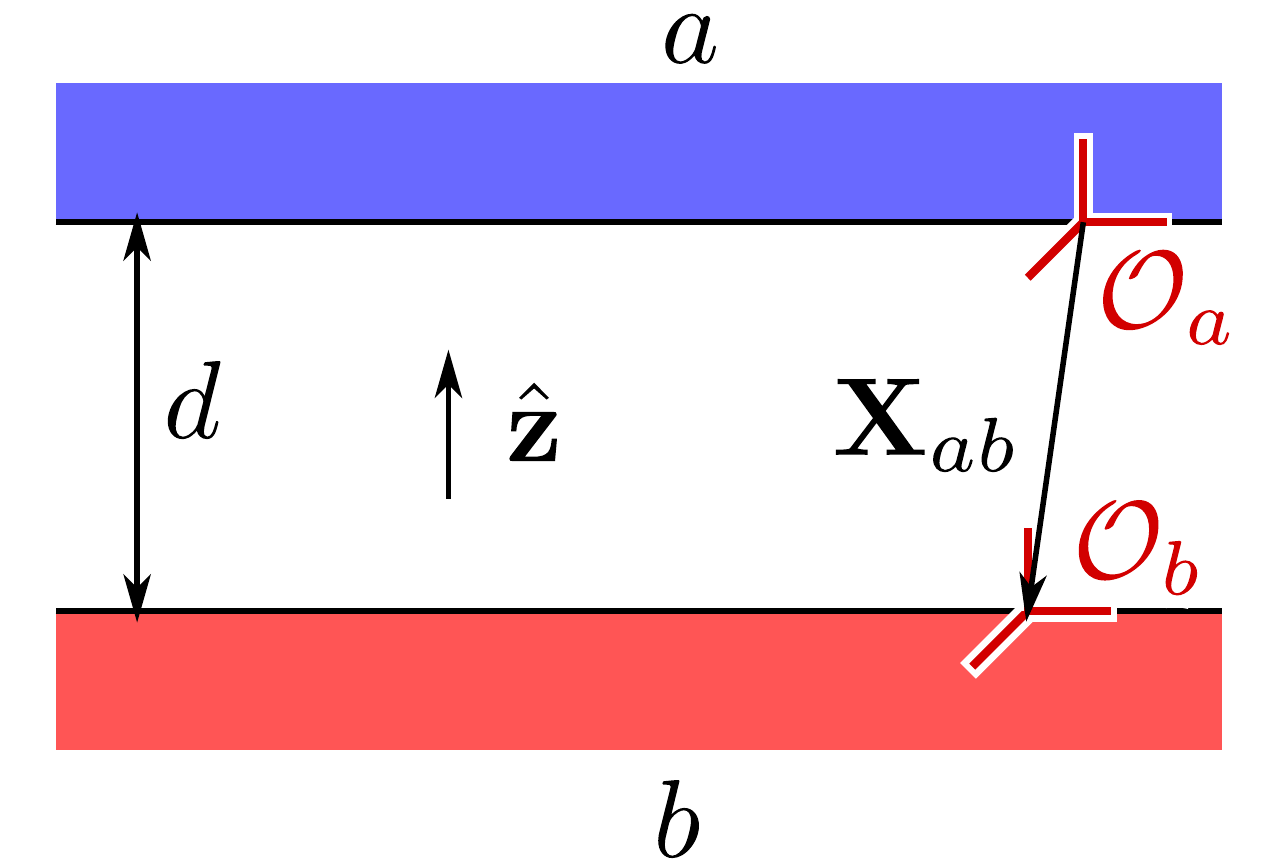}
\caption{(Color online) The upper infinite half space $a$ is located a
distance $d$ above the half space $b$. This is the original
configuration considered by Lifshitz. Each half space has its own
uniform electric permittivity $\epsilon_i(ic\kappa)$ and magnetic
permeability $\mu_i(ic\kappa)$.  We note that our calculation holds
even if the two origins $\mathcal{O}_a$ and $\mathcal{O}_b$ are
displaced horizontally from one another, as shown here.
}
\label{fig:Lifshitz}
\end{figure}

Next we consider two semi-infinite half-spaces with uniform electric
and magnetic susceptibility, as depicted in \reffig{Lifshitz}
\cite{Lifshitz55,Lifshitz56,Lifshitz57,Dzyaloshinskii61,Lifshitz80}.
We choose a plane wave basis oriented along the $\hat{\vecz}$ axis.

We decompose the scattering amplitude into magnetic (transverse
electric) modes $\M$ and electric (transverse magnetic) modes $\N$.

For the upper object $a$ the scattering solution is \cite{Jackson98}
\be
\begin{split}
\E(\kappa,\vecx) & = \Mout_{\veckpe}(\kappa,\vecx) + 
\int \frac{L^2 d\veckpe'}{(2\pi)^2}
\left[
\Mr_{\veckpe'}(\kappa,\vecx) \Toutout_{a,\veckpe' M,\veckpe M} +
\Nr_{\veckpe'}(\kappa,\vecx) \Toutout_{a,\veckpe' E,\veckpe M}
\right], \\
\E(\kappa,\vecx) & = \Nout_{\veckpe}(\kappa,\vecx) +
\int \frac{L^2 d^2\veckpe'}{(2\pi)^2}
\left[
\Mr_{\veckpe'}(\kappa,\vecx) \Toutout_{a,\veckpe' M,\veckpe E} +
\Nr_{\veckpe'}(\kappa,\vecx) \Toutout_{a,\veckpe' E,\veckpe E}
\right],
\end{split}
\labeleqn{scattplate}
\ee
Here $L$ is the length of each side of the plates,
$\veckpe$ is the momentum perpendicular to the $\hatz$ direction, and
the subscripts $M$ and $E$ on the scattering amplitudes denote the
magnetic and electric polarizations respectively. We consider the
limit $L\to\infty$. The scattering amplitudes are given by
\be
\begin{split}
\Toutout_{a,\veckpe' E,\veckpe M} & = \Toutout_{a,\veckpe' M,\veckpe E}
= 0 \, , \\
\Toutout_{a,\veckpe' M,\veckpe M} & =
\tfrac{(2\pi)^2}{L^2}\delta^{(2)}(\veckpe-\veckpe') \,
r_a^M\left(ic\kappa,\sqrt{1+\veckpe^2/\kappa^2}^{-1}\right) \, ,\\
\Toutout_{a,\veckpe' E,\veckpe E} & =
\tfrac{(2\pi)^2}{L^2}\delta^{(2)}(\veckpe-\veckpe') \,
r_a^E\left(ic\kappa,\sqrt{1+\veckpe^2/\kappa^2}^{-1}\right) \, ,
\end{split}
\labeleqn{Tplate}
\ee
in terms of the Fresnel coefficients
\be
\begin{split}
r_a^M(ic\kappa,x) & =
\frac
{\mu_a(ic\kappa) - \sqrt{1+(n^2_a(ic\kappa)-1)x^2}}
{\mu_a(ic\kappa) + \sqrt{1+(n^2_a(ic\kappa)-1)x^2}} \, , \\
r_a^E(ic\kappa,x) & =
\frac
{\epsilon_a(ic\kappa) - \sqrt{1+(n_a^2(ic\kappa)-1)x^2}}
{\epsilon_a(ic\kappa) + \sqrt{1+(n_a^2(ic\kappa)-1)x^2}}.
\end{split}
\labeleqn{Fresnel}
\ee
Here, $n_a$ is the index of refraction,
$n_a(ic\kappa)=\sqrt{\epsilon_a(ic\kappa) \mu_a(ic\kappa)}$.  In the
literature the Fresnel coefficients are also sometimes labeled with
$s$ instead of $M$ and $p$ in place of $E$.

The lower object $b$ has the same scattering properties. The relevant
scattering equation is the same as in \refeqn{scattplate}, with ``reg''
and ``out'' exchanged and $\Toutout_a$ replaced by
$\Tregreg_b$, which is obtained from $\Toutout_a$ simply by
substituting the permittivity $\epsilon_b(ic\kappa)$ and permeability
$\mu_b(ic\kappa)$ in place of those of object~$a$.

Using the appropriate $\X$ submatrices as specified in \refeqn{G0cases}
and the corresponding submatrices of $\F$, the energy \refeq{Elogdet2}
for two objects can be expressed as
\be
\begin{split}
\mathcal{E}
& = \frac{\hbar c}{2\pi} \int_0^\infty d\kappa \log \det
\left(
\bI - \Toutout_a \mathcal{W}^{ba} \Tregreg_b \mathcal{V}^{ba}
\right).
\end{split}
\labeleqn{ELifshitz}
\ee

Since the matrix in the determinant is diagonal in $\veckpe$, the
determinant factors into a product of determinants, each with fixed
$\veckpe$.  The logarithm of the product is
then given by an integral over the two-dimensional space of
$\veckpe$. Since the integrand is invariant under rotations in
$\veckpe$, we can write this integral in polar coordinates as
\be
\mathcal{E} = \frac{\hbar c}{2\pi}\int_0^\infty d\kappa \int_0^\infty
\frac{L^2}{2\pi} k_\perp dk_\perp \log\prod_{i=E,M}
\left(1 - r_a^i r_b^i e^{-2 \kappa d \sqrt{1+\veckpe^2/\kappa^2}}\right),
\ee
where $k_\perp = |\veckpe|$.

After a change of variable $p=\sqrt{1+\veckpe^2/\kappa^2}$ we obtain
the Lifshitz formula for the energy,
\be
\mathcal{E} = \frac{\hbar c L^2}{(2\pi)^2}
\int_0^\infty \kappa^2 d\kappa
\int_1^\infty p dp \log
\left[
\left(1 - r_a^M r_b^M e^{-2 \kappa p d}\right)
\left(1 - r_a^E r_b^E e^{-2 \kappa p d}\right)
\right].
\ee

\subsection{Two cylinders}

We now rederive the Casimir energy for two perfectly conducting, 
infinitely long cylinders.  The result for one cylinder inside the other has
been presented in Ref. \cite{Dalvit06} and the result for both outside
each other was presented in Refs. \cite{Rahi08-2,Rahi08-1}.

\begin{figure}[ht]
\includegraphics[scale=0.45]{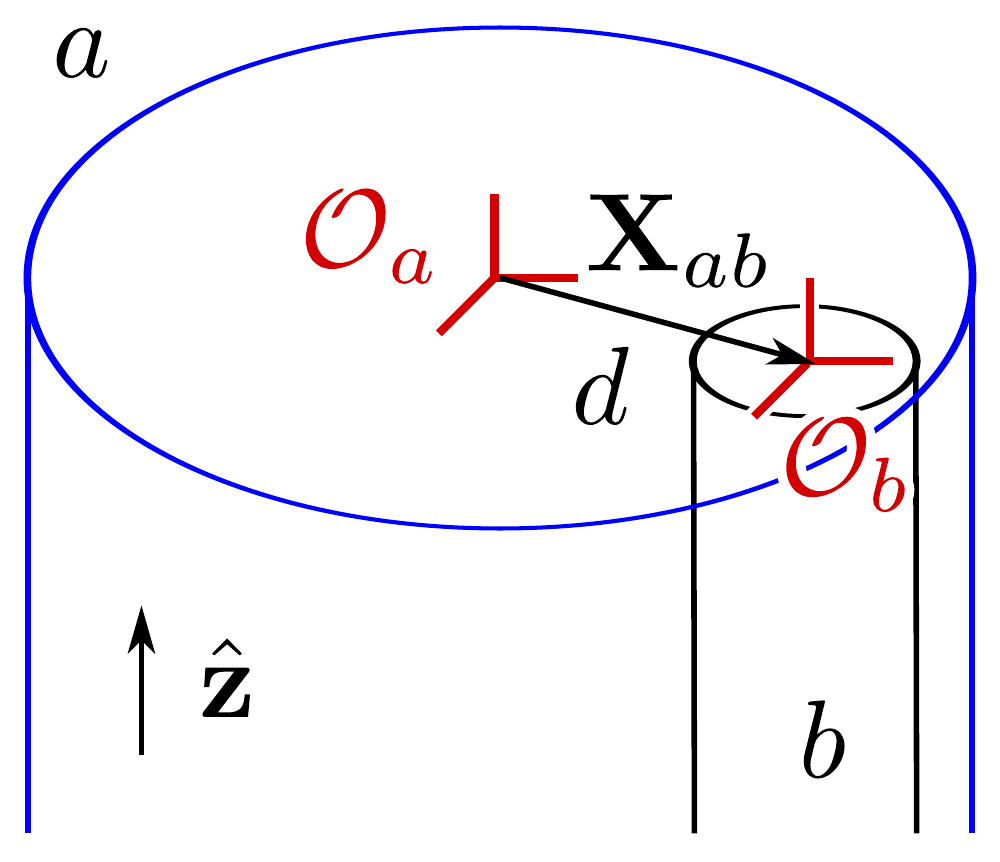}
\hspace{1.5cm}
\includegraphics[scale=0.45]{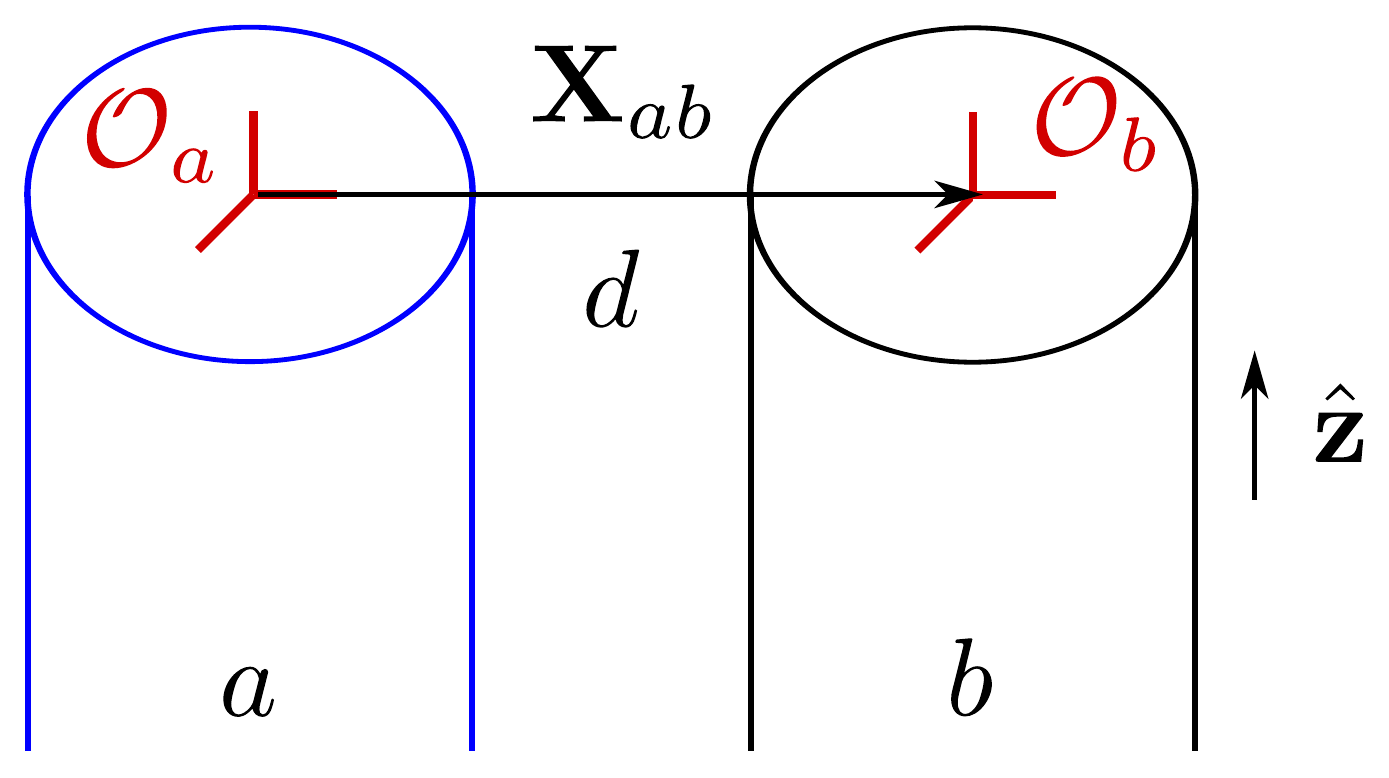}
\caption{(Color online) Two perfectly conducting infinite cylinders
with radii $R_a$ and $R_b$ are separated by a center-to-center
distance $d$. They can be outside one another, or one may be inside
the other.}
\label{fig:twocyls}
\end{figure}

For scattering from outside cylinder $a$, we have the
scattering solutions
\be
\begin{split}
\E(\kappa,\vecx) & = \Mr_{k_z,n}(\kappa,\vecx) + \int \frac{L dk_z'}{2\pi}
 \sum_{n'}
\left[\Mout_{k_z',n'}(\kappa,\vecx) \Tregreg_{a,k_z' n' M,k_z n M} +
\Nout_{k_z',n'}(\kappa,\vecx) \Tregreg_{a,k_z' n' E,k_z n M}\right], \\
\E(\kappa,\vecx) & = \Nr_{k_z,n}(\kappa,\vecx) + \int \frac{L
 dk_z'}{2\pi} \sum_{n'}
\left[\Mout_{k_z',n'}(\kappa,\vecx) \Tregreg_{a,k_z' n' M,k_z n E} +
\Nout_{k_z',n'}(\kappa,\vecx) \Tregreg_{a,k_z' n' E,k_z n E}\right],
\end{split}
\labeleqn{scattcyloutside}
\ee
with boundary conditions $\E^\parallel=0$ and $\B^\perp=0$ on
the cylinder surface. $L$ is the length of the cylinders, and we are
considering the limit $L\to\infty$. We have
\be
\begin{split}
\Tregreg_{a,k_z' n' E,k_z n M} & =
\Tregreg_{a,k_z' n' M,k_z n E} = 0, \\
\Tregreg_{a,k_z' n' M,k_z n M} & =
-\tfrac{2\pi}{L}\delta(k_z-k_z')\delta_{n,n'}
\frac{I'_n \left( R_a p\right)}{K'_n\left(R_a p\right)}\, , \\
\Tregreg_{a,k_z' n' E,k_z n E} & =
-\tfrac{2\pi}{L}\delta(k_z-k_z')\delta_{n,n'}
\frac{I_n \left( R_a p\right)}{K_n\left(R_a p\right)},
\end{split}
\ee
and analogous equations hold for scattering from cylinder $b$.
The energy in \refeqn{Elogdet2} is given in terms of
exterior scattering amplitudes only,
\be
\begin{split}
\mathcal{E}
& = \frac{\hbar c}{2\pi} \int_0^\infty d\kappa \log\det 
\left(\bI-\Tregreg_a \mathcal{U}^{ba} \Tregreg_b \mathcal{U}^{ab} \right).
\end{split}
\ee
The matrix inside the determinant is diagonal in $k_z$, so the
log-determinant over this index turns into an overall integral.  
A change of variable to polar coordinates converts the integrals over
$\kappa$ and $k_z$ to a single integral over $p=\sqrt{k_z^2+\kappa^2}$,
yielding
\be
\mathcal{E} = \frac{\hbar c L}{4\pi}\int_0^\infty pdp
\left(\log\det \Y^M + \log\det \Y^E\right),
\ee
where
\be
\begin{split}
\Y^M_{n,n''} & = \delta_{n,n''} - \sum_{n'}
\frac{I'_n(p R_a)}{K'_n(p R_a)} K_{n+n'}(pd)
\frac{I'_{n'}(p R_b)}{K'_{n'}(p R_b)} K_{n'+n''}(pd) \\
\Y^E_{n,n''} & = \delta_{n,n''} - \sum_{n'}
\frac{I_n(p R_a)}{K_n(p R_a)} K_{n+n'}(pd)
\frac{I_{n'}(p R_b)}{K_{n'}(p R_b)} K_{n'+n''}(pd).
\end{split}
\ee

For scattering from inside cylinder $a$ we have the scattering solutions
\be
\begin{split}
\E(\kappa,\vecx) & = \Mout_{k_z,n}(\kappa,\vecx) + \int \frac{L
dk_z'}{2\pi} \sum_{n'}
\left[\Mr_{k_z',n'}(\kappa,\vecx) \Toutout_{a,k_z' n' M,k_z n M} +
\Nr_{k_z',n'}(\kappa,\vecx) \Toutout_{a,k_z' n' E,k_z n M}\right], \\
\E(\kappa,\vecx) & = \Nout_{k_z,n}(\kappa,\vecx) + \int \frac{L
dk_z'}{2\pi} \sum_{n'}
\left[\Mr_{k_z',n'}(\kappa,\vecx) \Toutout_{a,k_z' n' M,k_z n E} +
\Nr_{k_z',n'}(\kappa,\vecx) \Toutout_{a,k_z' n' E,k_z n E}\right],
\end{split}
\ee
yielding
\be
\begin{split}
\Toutout_{a,k_z' n' E,k_z n M} & =
\Toutout_{a,k_z' n' M,k_z n E} = 0 \, ,\\
\Toutout_{a,k_z' n' M,k_z n M} & =
-\tfrac{2\pi}{L}\delta(k_z-k_z')\delta_{n,n'}
\frac{K'_n \left( R_a p\right)}{I'_n\left(R_a p\right)} \, ,\\
\Toutout_{a,k_z' n' E,k_z n E} & =
-\tfrac{2\pi}{L}\delta(k_z-k_z')\delta_{n,n'}
\frac{K_n \left( R_a p\right)}{I_n\left(R_a p\right)}.
\end{split}
\ee
We note that the inside scattering amplitude matrix is the the
inverse of the corresponding outside result.  The energy, expressed in
\refeqn{Elogdet2}, now becomes
\be
\begin{split}
\mathcal{E}
& = \frac{\hbar c}{2\pi} \int_0^\infty d\kappa \log \det
\left(
\bI - \Toutout_a \mathcal{W}^{ba} \Tregreg_b \mathcal{V}^{ba}
\right),
\end{split}
\ee
which contains the appropriate scattering amplitudes for the inside
problem, $\Toutout_a$ for interior scattering of object $a$ and
$\Tregreg_b$ for exterior scattering of object $b$.  Using the same
simplifications as in the outside case, we have
\be
\mathcal{E}  = \frac{\hbar c L}{4\pi}\int_0^\infty pdp
\left( \log\det \Y^M + \log\det \Y^E\right),
\ee
where
\be
\begin{split}
\Y^M_{n,n''} & = 
\delta_{n,n''}-
\sum_{n'}\frac{K'_n(p R_a)}{I'_n(p R_a)}I_{n+n'}(pd)\frac{I'_{n'}(p
R_b)}{K'_{n'}(p R_b)}I_{n'+n''}(pd)\, , \\
\Y^E_{n,n''} & = 
\delta_{n,n''}-
\sum_{n'}\frac{K_n(p R_a)}{I_n(p R_a)}I_{n+n'}(pd)\frac{I_{n'}(p
R_b)}{K_{n'}(p R_b)}I_{n'+n''}(pd).
\end{split}
\ee

\subsection{Sphere and plate}

\begin{figure}[ht]
\includegraphics[scale=0.45]{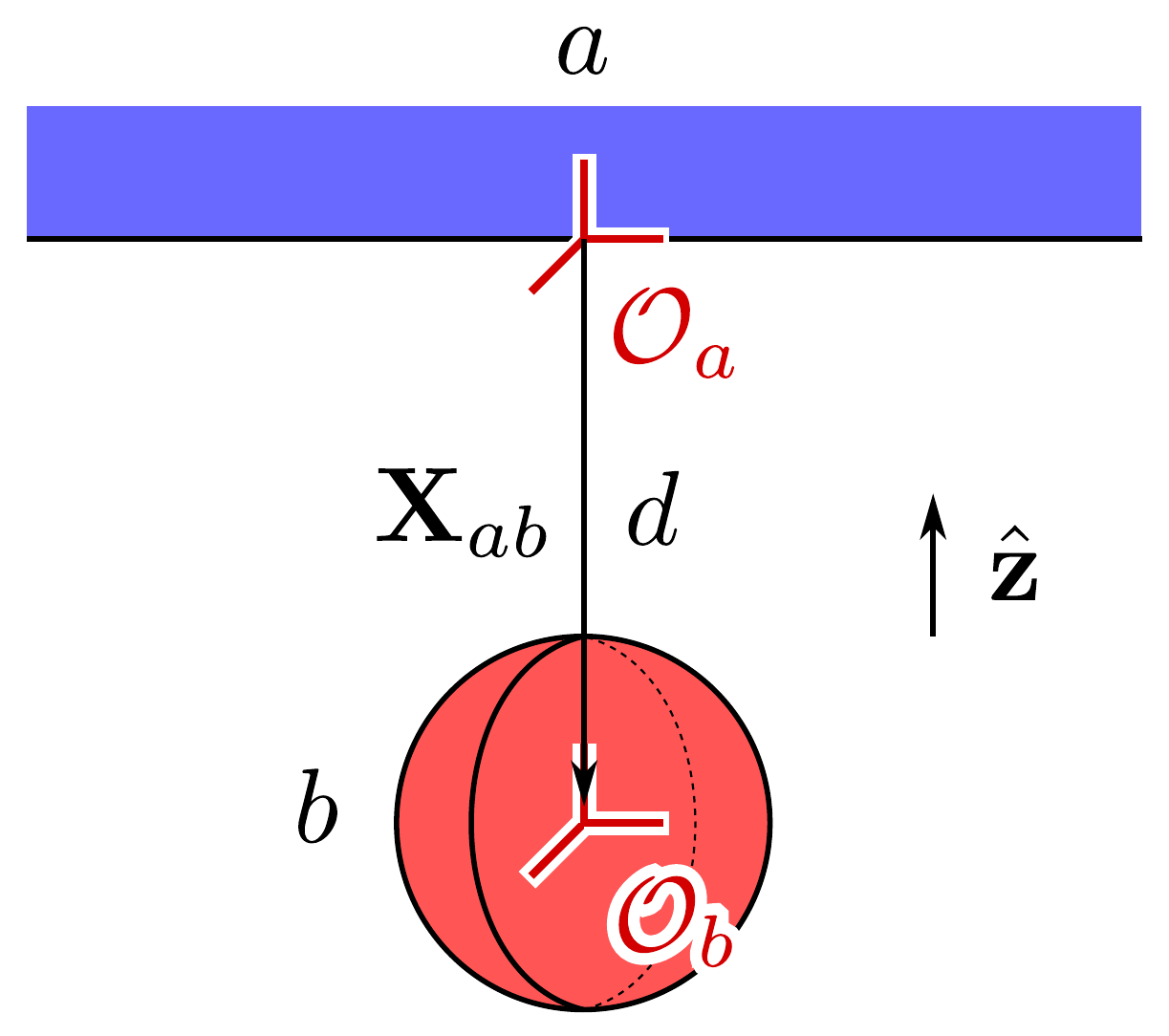}
\caption{(Color online) A sphere $b$ of radius $R$ is located opposite
a plate $a$, separated by a center-to-surface distance $d$.}
\label{fig:platesphere}
\end{figure}

In this section we investigate the Casimir interaction of an infinitely thick
plate $a$ opposite a sphere $b$, each with frequency-dependent
permittivity and permeability.  The geometry is depicted in
\reffig{platesphere}.

The scattering amplitude for the plate is easy to express in the plane wave
basis using \refeqn{Tplate}.  We can apply our result from the plane
geometry, \refeqn{ELifshitz}, changing only  $\Tregreg_b$, which now
becomes the scattering amplitude for vector plane wave functions outside a
sphere. To express the scattering amplitude of the sphere in
the spherical vector wave basis, we use Eqs. \refeq{convMNsphereplane}
and \refeq{convDsphereplane} and obtain
\be
\begin{split}
\Tregreg_{b,\veckpe P,\veckpe' P'}
& =
(-1) C_{\veckpe P}(\kappa)
\langle\E^\text{reg}_{ \veckpe P}(\kappa) | \T_b |
\E^\text{reg}_{\veckpe' P'}(\kappa) \rangle \\
& = \sum_{lmQ,l'm'Q'}
(-1) \frac{C_{\veckpe P}(\kappa)}{C_{Q}(\kappa)}
D^\dagger_{\veckpe P, l m Q}
C_{Q}(\kappa)
\langle\E^\text{reg}_{l m Q}(\kappa) | \T_b |
\E^\text{reg}_{l' m' Q'}(\kappa) \rangle
D_{l' m' Q', \veckpe' P'} \\
& = \sum_{lmQ,l'm'Q'}
\frac{C_{\veckpe P}(\kappa)}{C_{Q}(\kappa)}
D^\dagger_{\veckpe P, l m Q}
\Tregreg_{b,lmQ,l'm'Q'}
D_{l' m' Q', \veckpe' P'},
\end{split}
\labeleqn{Tsphereinplate}
\ee
where $P$ and $P'$ label the polarizations in the
planar basis and $Q$ and $Q'$ label the polarizations in the spherical
basis.  The normalization factors 
$C_{\veckpe P}(\kappa)$ and $C_{Q}(\kappa)$, defined below
\refeqn{tG0plane} and \refeqn{tG0sphere},
arise from the definition of the scattering amplitude (see, for
example, \refeqn{Temii}). For a sphere with uniform permittivity and
permeability, we compute the scattering amplitude by solving
\refeqn{Temii} in the spherical vector wave basis, which yields
\be
\begin{split}
\Tregreg_{b,lmE,l'm'M} & = \Tregreg_{b,lmM,l'm'E} = 0 \, ,\\
\Tregreg_{b,lmM,l'm'M} & =
-\delta_{l,l'}\delta_{m,m'}
\frac{i_l(\kappa R) \partial_R (R i_l(n_b\kappa R))
- \mu_b \partial_R (R i_l(\kappa R)) i_l(n_b\kappa R)}
{k_l(\kappa R) \partial_R (R i_l(n_b\kappa R))
- \mu_b \partial_R (R k_l(\kappa R)) i_l(n_b\kappa R)} \, ,\\
\Tregreg_{b,lmE,l'm'E} & =
-\delta_{l,l'}\delta_{m,m'}
\frac{i_l(\kappa R) \partial_R (R i_l(n_b\kappa R))
- \epsilon_b \partial_R (R i_l(\kappa R)) i_l(n_b\kappa R)}
{k_l(\kappa R) \partial_R (R i_l(n_b\kappa R))
- \epsilon_b \partial_R (R k_l(\kappa R)) i_l(n_b\kappa R)},
\end{split}
\ee
where $n_b$ is the index of refraction,
$n_b(ic\kappa)=\sqrt{\epsilon_b(ic\kappa) \mu_b(ic\kappa)}$.  The
modified spherical Bessel functions $i_l$ and $k_l$ are defined in
Appendix~\ref{sec:Greenexp}.

Plugging into \refeqn{ELifshitz} and using
$\det(\tI+\tA\tB)=\det(\tI+\tB\tA)$, the energy simplifies to
\be
\mathcal{E} = \frac{\hbar c}{2\pi}\int_0^\infty d\kappa \log\det
\left(\bI-\Y\right),
\labeleqn{Eplatesphere}
\ee
where
\be
\begin{split}
\Y_{lmP,l'm'P'} & =  \delta_{m,m'} \, 
\Tregreg_{b,lmP,lmP} \\
& \times \int_0^\infty\frac{k_\perp dk_\perp}{2\pi}
\tfrac{e^{-2d\sqrt{\veckpe^2+\kappa^2}}}{2\kappa\sqrt{\veckpe^2+\kappa^2}} \\
& \times \sum_Q
D_{lm P,\veckpe Q} \,
r^{Q}_a\left(ic\kappa,\sqrt{1+\veckpe^2/\kappa^2}^{-1}\right) \, 
D^\dagger_{\veckpe Q,l'm P'}
(2 \delta_{Q,P'}-1).
\end{split}
\ee
Here $k_\perp= |\veckpe|$ and $r_a^Q$,
defined in \refeqn{Fresnel}, is the Fresnel coefficient for reflection
of a wave with polarization $Q$. The ratio of $C_{\veckpe P}(\kappa)$
to $C_{Q}(\kappa)$ in \refeqn{Tsphereinplate} has opposite signs
depending on whether $P$ and $Q$ represent the same polarization 
or the opposite polarization, which we have implemented through the
term $(2 \delta_{P,Q}-1)$. The integration over all angles of
$\veckpe$ has already been carried out, which makes $\Y$ diagonal in
$m$ and $m'$.  (Although $D_{lmp,\veckpe Q}$ seems to depend on the angle of
$\veckpe$, the multiplication with its Hermitian conjugate cancels this
dependence.)

\begin{figure}[ht]
\includegraphics[scale=0.65]{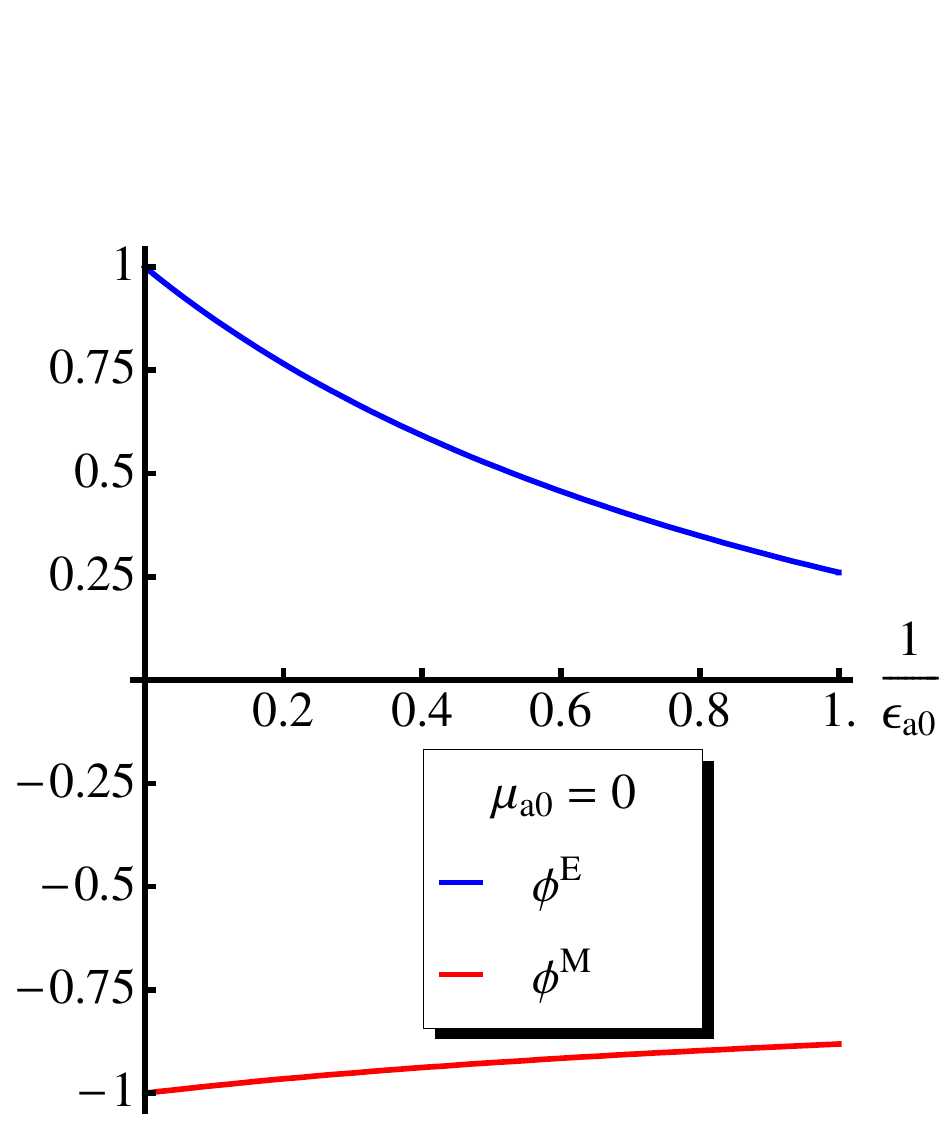}
\hspace{1.5cm}
\includegraphics[scale=0.65]{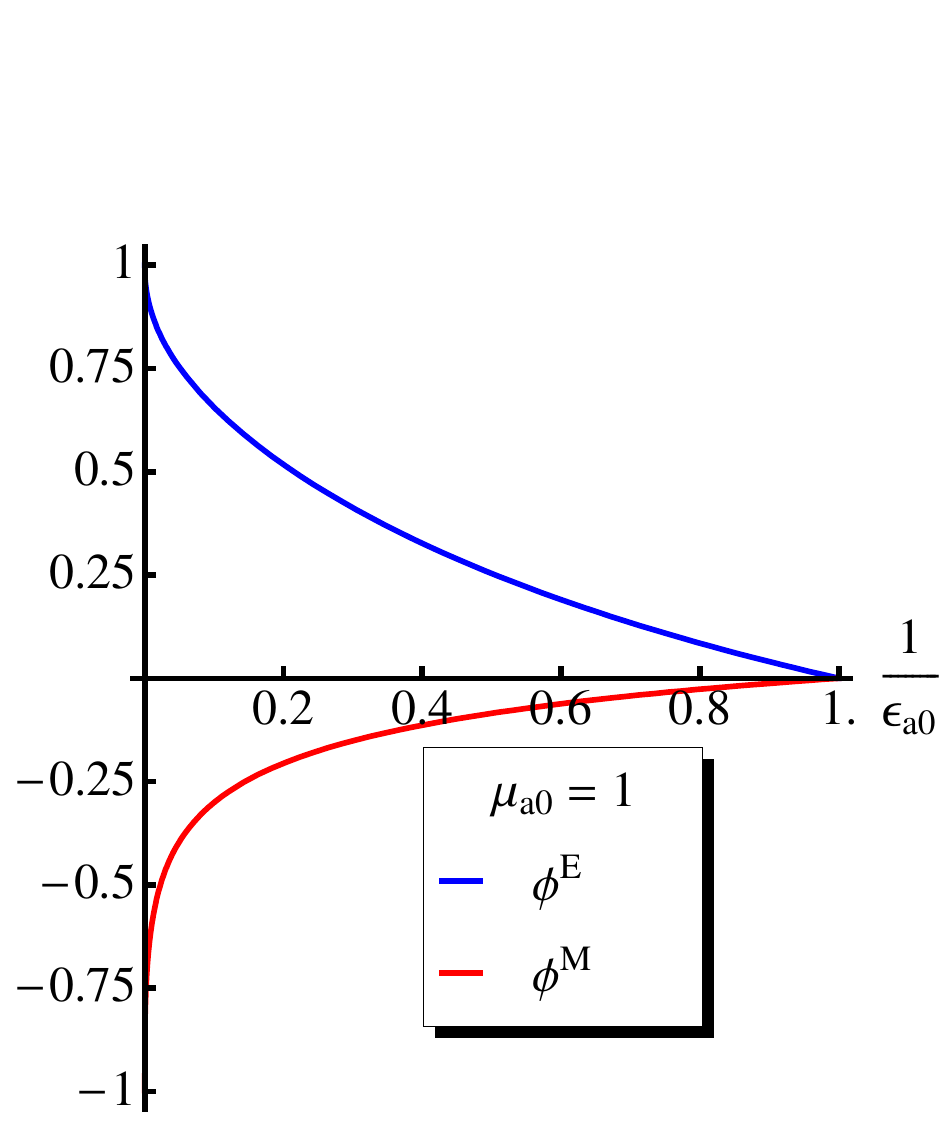}
\caption{(Color online) Plots of $\phi^E$ (blue, positive) and
$\phi^M$ (red, negative) as functions of $1/\epsilon_{a0}$ for fixed
$\mu_{a0}=0$ (left) or fixed $\mu_{a0}=1$ (right). For $\mu_{a0}=1$ the two functions $\phi^E$ and $\phi^M$
approach $1$ rather slowly from the right (perfect metal limit). So, for
comparison with experiments, it may not be justified to use the
perfect metal limit $\epsilon_a\to\infty$ of the plate to compute the
Casimir energy.}
\label{fig:platesphereasym}
\end{figure}

To leading order for large $d/R$, the $l=1$ components of the sphere's
scattering amplitude and the $\kappa\to 0$ limit of the permittivities
and permeabilities contribute. The scattering amplitude
can be expanded to lowest order in terms of the sphere's electric and
magnetic polarizabilities,
$\Tregreg_{b,1mM,1mM}\to \frac{2}{3}\alpha^M
\kappa^3$ and $\Tregreg_{b,1mE,1mE}\to \frac{2}{3}\alpha^E \kappa^3$,
where the polarizabilities $\alpha^M=\frac{\mu_{b0}-1}{\mu_{b0}+2}R^3$ and
$\alpha^E=\frac{\epsilon_{b0}-1}{\epsilon_{b0}+2}R^3$ are given in terms
of the zero frequency permittivity $\epsilon_{b0}=\epsilon_{b}(0)$ and
permeability $\mu_{b0}=\mu_{b}(0)$. To leading order the energy is given by
\be
\mathcal{E} = -\frac{3\hbar c}{8\pi d^4} (\alpha^M \phi^M+\alpha^E \phi^E),
\labeleqn{Eplatesphereasym}
\ee
where
\be
\begin{split}
\phi^M & = \int_0^1 dx \, \left[\left(1-\frac{x^2}2\right)
r^M_a(0,x) 
-\frac{x^2}{2}
r^E_a(0,x)
\right] \, ,
\\
\phi^E & = \int_0^1 dx \, \left[\left(1-\frac{x^2}{2}\right)
r^E_a(0,x)
-\frac{x^2}{2}
r^M_a(0,x)
\right]
\end{split}
\ee
can be expressed in terms of elementary functions, but the
expressions are too complicated to be worth reproducing here. The two
functions are plotted in \reffig{platesphereasym}.

The expression for the energy in \refeqn{Eplatesphereasym} agrees with
the results in Ref. \cite{Emig08-1} for a perfect metal plate
$\epsilon_a\to\infty$ and a sphere with general $\epsilon_b$.  It also
agrees with the results in Ref. \cite{Maia_Neto08} for a perfect
metal plate and a perfect metal sphere, $\epsilon_a\to\infty$, and
$\epsilon_b\to\infty$ .  Both of these works arrive at similar general
expressions for the energy to what we have found here;
Ref. \cite{Emig08-1} combines scattering theory techniques we have
used here with the method of images, while Ref. \cite{Maia_Neto08}
uses Wigner rotation matrices.

In the calculations of Refs. \cite{Emig08-1} and \cite{Maia_Neto08},
when $\epsilon \to \infty$ the corresponding $\mu$ is set to zero to
reproduce perfect metal boundary conditions within a
low-frequency expansion.  (Ref. \cite{Dzyaloshinskii61} contains the
asymptotic Casimir energy formula \refeqn{Eplatesphereasym} in the
case where the magnetic permeabilities are set equal to one.)
As the plots in \reffig{platesphereasym} show, however, the perfect
reflectivity limit of the plate is approached slowly with
increasing $\epsilon_{a}$. To compare with experiments it is
thus important to compute the energy \refeq{Eplatesphere}
using the actual permittivities and permeabilities of the material
instead of perfect metal limits.

\subsection{Cylinder and plate}

\begin{figure}[ht]
\includegraphics[scale=0.45]{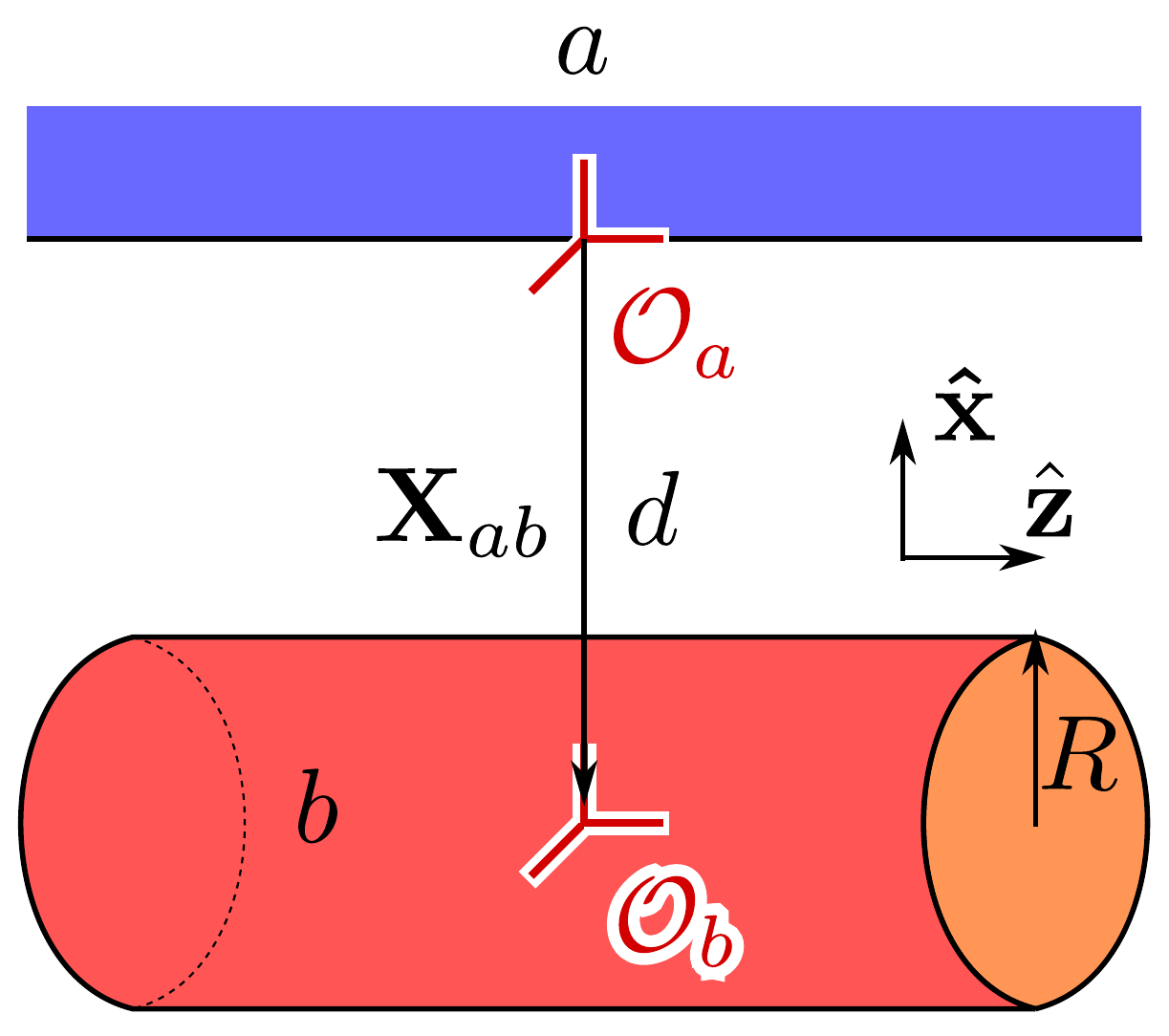}
\caption{(Color online) A cylinder $b$ of radius $R$ is located opposite
a plate $a$, separated by a center-to-surface distance $d$.}
\label{fig:platecyl}
\end{figure}

As a final example, we investigate the Casimir interaction energy
between an infinitely thick plate $a$ opposite a cylinder $b$, each
with frequency-dependent permittivity and permeability
(\reffig{platecyl}). We will focus on presenting asymptotic
($d/R\to\infty$) results here, but the derivation is straightforward
to extend to intermediate separations, for which the evaluation of the
final expression can be performed easily on a computer. We choose the
$\hatz$ axis as the axis of symmetry of the cylinder and let $\veckpe$
denote the vector $(k_y,k_z)$.

Just as in the sphere-plate case, it is most convenient to express the
scattering amplitude of the cylinder $b$ in a plane wave basis by
\be
\begin{split}
\Tregreg_{b,\veckpe P,\veckpe' P'}
& =
(-1) C_{\veckpe P}(\kappa)
\langle\E^\text{reg}_{ \veckpe P}(\kappa) | \T_b |
\E^\text{reg}_{\veckpe' P'}(\kappa) \rangle \\
& = \sum_{nQ,n'Q'}
(-1) \frac{C_{\veckpe P}(\kappa)}{C_{Q}}
D^\dagger_{\veckpe P, k_z n Q}
C_{Q}
\langle\E^\text{reg}_{k_z n Q}(\kappa) | \T_b |
\E^\text{reg}_{k_z n' Q'}(\kappa) \rangle
D_{k_z n' Q', \veckpe' P'} \\
& = \sum_{nQ,n'Q'}
\frac{C_{\veckpe P}(\kappa)}{C_{Q}}
D^\dagger_{\veckpe P, k_z n Q}
\Tregreg_{b,k_z n Q,k_z n'Q'}
D_{k_z n' Q', \veckpe' P'},
\end{split}
\labeleqn{Tcylinplate}
\ee
where $C_{\veckpe P}(\kappa)$ and $C_{Q}$ are defined below
\refeqn{tG0plane} and \refeqn{tG0cyl}, respectively.

By solving \refeqn{Temii} in a cylindrical wave basis, it is
straightforward to find the scattering amplitude of the cylinder,
$\Tregreg_{b,k_z n Q, k'_z n' Q'}$.  For uniform permittivity and
permeability, the matrix elements are diagonal in $k_z$ and
the cylindrical wave index $n$, but not in TE and TM polarization.  The
expressions are somewhat complicated; since we are presenting
asymptotic results here, we only need the small-radius expansion,
\be
\begin{split}
\Tregreg_{b,k_z n P, k_z' n' P'} & = \tfrac{2 \pi}{L} \delta(k_z-k_z')
\delta_{n,n'} f_{k_z n P P'} + O(R^4) \, ,\\
f_{k_z0MM} & = 
\tfrac{1}{2} (\kappa^2+k_z^2) R^2
\, (1-\mu_b) \, , \qquad
f_{k_z0EE} = 
\tfrac{1}{2} (\kappa^2+k_z^2) R^2
\, (1-\epsilon_b) \, ,\\
f_{k_z\pm1MM} & =
\tfrac{k_z^2(1+\epsilon_b)(1-\mu_b)-\kappa^2(1-\epsilon_b)(1+\mu_b)}
{2(1+\epsilon_b)(1+\mu_b)} R^2 \, , \qquad
f_{k_z\pm1EE} =
\tfrac{k_z^2(1-\epsilon_b)(1+\mu_b)-\kappa^2(1+\epsilon_b)(1-\mu_b)}
{2(1+\epsilon_b)(1+\mu_b)} R^2 \, ,\\
f_{k_z1ME} & = f_{k_z-1EM} =
\tfrac{\kappa k_z(\epsilon_b \mu_b-1)}{(1+\epsilon_b)(1+\mu_b)} R^2
\,, \qquad
f_{k_z1EM} = f_{k_z-1ME} = - f_{k_z1ME}.
\end{split}
\labeleqn{fcyl}
\ee
All other matrix elements ($|n|>1$) contribute at higher order in
$R$. It is assumed here that $\epsilon_b(ic\kappa)$ is finite. In the
infinite conductivity limit ($\epsilon_b\to\infty$) only one of these
scattering amplitudes contributes; this case is discussed below.

We next plug into \refeqn{ELifshitz}.  As in the case of two cylinders,
the matrix inside the determinant is diagonal in $k_z$, so the
log-determinant over this index turns into an integral.  We obtain for
the Casimir energy
\be
\mathcal{E} =
\frac{\hbar c L}{4 \pi^2}
\int_0^\infty d\kappa \int_{-\infty}^\infty dk_z \,
\log\det\left(\bI-\Y\right),
\labeleqn{Ecylplate}
\ee
where
\be
\begin{split}
\Y_{k_z n P,n' P'} & = 
\sum_{P'',Q}
f_{k_z n P P''} \int_{-\infty}^\infty dk_y
\tfrac{e^{-2d\sqrt{\veckpe^2+\kappa^2}}}{2 \sqrt{\veckpe^2+\kappa^2}} \\
& \times
D_{n k_z P'',\veckpe Q} \,
r^Q_{a}\left(ic\kappa,\sqrt{1+\veckpe^2/\kappa^2}^{-1}\right)
D^\dagger_{\veckpe Q, n' k_z P'}
(1-2\delta_{Q,P'}).
\end{split}
\ee

To find the interaction energy at separations outside of the
asymptotic limit, $f_{k_z n P P'}$ must be replaced by the appropriate
scattering amplitude expressions for all $n$, valid to all orders
in $R$. Expanding the $\log\det$ in \refeqn{Ecylplate} to first order
in $\mathcal{N}$, we obtain for the interaction in the large distance
limit $d/R\to\infty$,
\be
\mathcal{E} = -\frac{3 \hbar c L R^2}{128 \pi d^4}
\int_0^1 dx \frac{\epsilon_{b0}-1}{\epsilon_{b0}+1}
\left[(7 + \epsilon_{b0} - 4 x^2)r^E(0,x)-(3+\epsilon_{b0}) x^2
 r^M(0,x)\right],
\ee
if the zero-frequency magnetic permeability $\mu_{b0}$ of the cylinder
is set to one. If we do not set $\mu_{b0}$ equal to one, but instead
take the perfect reflectivity limit for the plate, we obtain
\be
\mathcal{E} = -\frac{\hbar c L R^2}{32 \pi d^4}
\frac{(\epsilon_{b0}-\mu_{b0})(
9+\epsilon_{b0}+\mu_{b0}+\epsilon_{b0}\mu_{b0})}
{(1+\epsilon_{b0})(1+\mu_{b0})}.
\ee

\begin{figure}[ht]
\includegraphics[scale=0.65]{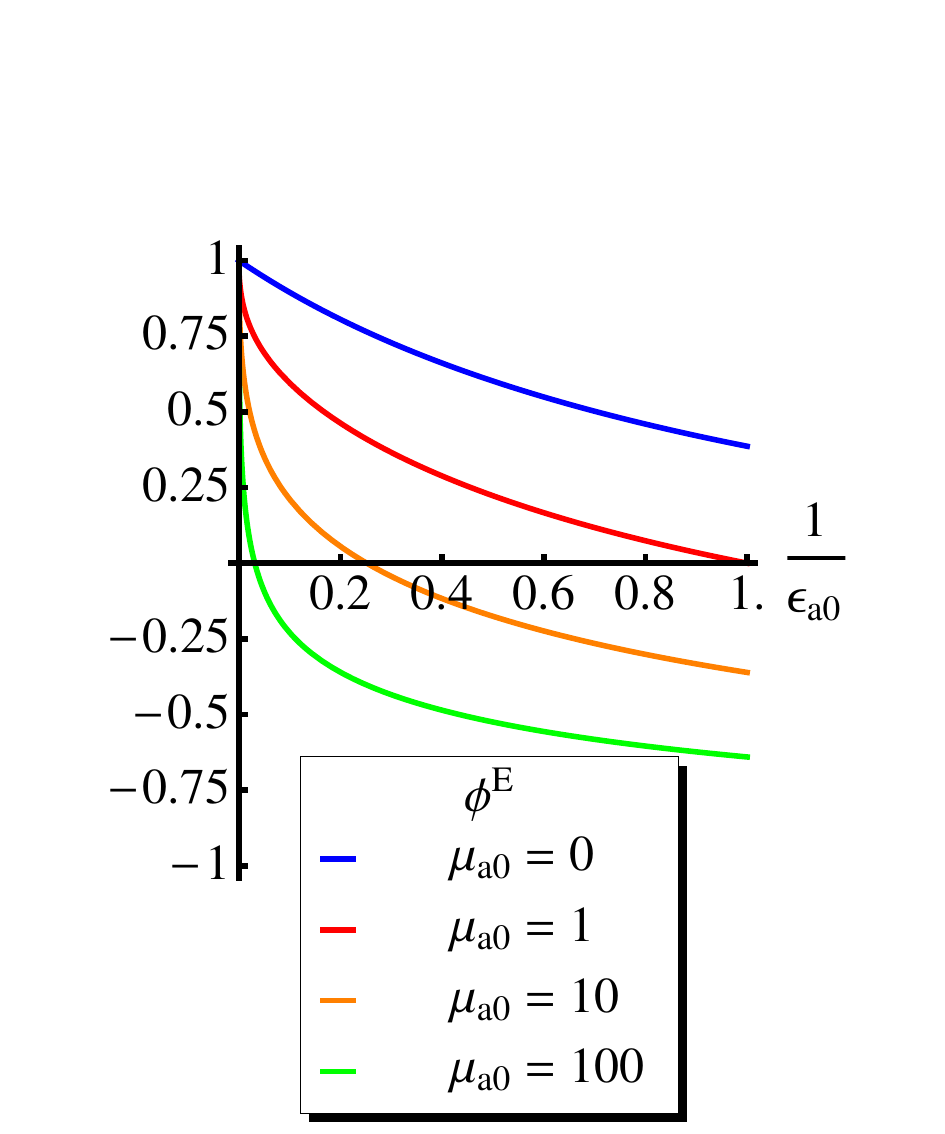}
\caption{(Color online) Plots of $\phi^E$ versus $1/\epsilon_{a0}$ for
fixed values of $\mu_{a0}$. The perfect metal limit ($\phi^E=1$) is
approached slowly for large $\mu_{a0}$, as in the case of a sphere
opposite a plate. For large $\mu_{a0}$ the interaction becomes
repulsive, which is expected given similar results for two infinite
plates \cite{Kenneth02}.}
\label{fig:platecylasym}
\end{figure}

Finally, if we let $\epsilon_b$ be infinite from the beginning (the
perfect metal limit for the cylinder), only the $n=0$ TM mode of the
scattering amplitude, $\Tregreg_{b,k_z 0 E, k_z' 0 E} =
\tfrac{2\pi}{L}\delta(k_z-k_z')\frac{1}{\log{R/d}} +
O(\log^{-2}(R/d))$, contributes at lowest order; the previous
expansions of the cylinder's scattering amplitude in \refeqn{fcyl} are
not valid.  For a plate with zero-frequency permittivity
$\epsilon_{a0}$ and permeability $\mu_{a0}$, we obtain for the Casimir
energy
\be
\mathcal{E} = \frac{\hbar c L}{16 \pi d^2 \log(R/d)} \phi^E \, ,
\ee
where
\be
\phi^E = \int_0^1 \frac{dx}{1+x}\left[r^E_a(0,x)-x r^M_a(0,x)\right].
\ee
In \reffig{platecylasym}, $\phi^E$ is plotted as a function of the
zero-frequency permittivity of the plate, $\epsilon_{a0}$, for various
zero-frequency permeability values, $\mu_{a0}$.

\begin{acknowledgments}
We thank S. Zaheer and the participants in the Kavli Institute
for Theoretical Physics (KITP) Workshop on the Theory and Practice of
Fluctuation-Induced Interactions for conversations and suggestions.

SJR would like to thank Simin and Reza Sharifzadeh for their limitless
generosity, which made this work possible.

This work was supported by the National Science Foundation (NSF)
through grants PHY05-55338 and PHY08-55426 (NG), DMR-08-03315 (SJR and
MK), and by the U. S. Department of Energy (DOE) under cooperative
research agreement \#DF-FC02-94ER40818 (RLJ).
Portions of this work were carried out at the KITP, which is supported
in part by NSF under grant PHY05-51164.
\end{acknowledgments}

\begin{appendix}

\section{Derivation of the macroscopic field theory}
\label{app:Derivation}

In this appendix we justify the starting point of our derivation, the
effective or ``macroscopic'' field theory in \refeqn{Lem}, in order to
clarify the causality properties of the permittivity
$\epsilon$ and permeability $\mu$ and incorporate the
dissipative properties of the materials naturally into our
formalism.  We also show that both non-local and non-isotropic
permittivity and permeability \emph{tensors} can be used.

In place of \refeqn{Lem},
we begin with the Lagrangian density of the free electromagnetic field
plus a coupled system of particles. To be concrete, we imagine that
the electromagnetic field couples, say, to electrons of charge $-e$ 
described by the Lagrangian density operator
\be
\hat{\mathcal{L}} = \thalf\left(\hat{\E}^2-\hat{\B}^2\right) +
i \hbar \hat{\psi}^\dagger \dot{\hat{\psi}} - \frac{\hbar^2}{2m}
\left[\left(\nabla-\frac{i e}{\hbar}
\hat{\A}\right)\hat{\psi}^\dagger\right]
\left[\left(\nabla+\frac{i e}{\hbar} \hat{\A}\right)\hat{\psi}\right],
\label{eq:L}
\ee
where $\hat{\psi}$ and $\hat{\psi}^\dagger$ are the fermion field
annihilation and creation operators, which are spinor functions of space
$\vecx$ and time $t$.  We implicitly sum over spins and
suppress the spin index and we continue to work in $A^0=0$
gauge.  The electrons' coupling to the lattice, their coupling to each
other, and their spin-magnetic field interaction can be explicitly
taken into account by adding, for example,
\be
\begin{split}
\hat{\mathcal{L}}_\text{lattice} & = - \hat{\psi}^\dagger(t,\vecx)
u(\vecx) \hat{\psi}(t,\vecx) \\
\hat{\mathcal{L}}_\text{int} & = - \thalf\int d\vecx' \,
\hat{\psi}^\dagger(t,\vecx) \hat{\psi}^\dagger(t,\vecx')
v(\vecx-\vecx') \hat{\psi}(t,\vecx') \hat{\psi}(t,\vecx) \\
\hat{\mathcal{L}}_\text{spin} & = \frac{e\hbar g_e}{2 m c}
\hat{\B}(\vecx) \cdot \hat{\psi}^\dagger(t,\vecx) 
\boldsymbol{\sigma} \hat{\psi}(t,\vecx).
\end{split}
\labeleqn{elecinter}
\ee
We imagine that such systems, confined to the various
regions of space, represent the objects whose Casimir
interactions we are calculating. Since the following procedure is quite
general, we are not limited to systems described by these particular
Lagrangians, but modifications to our approach may be needed in some
situations.

The electron-lattice and electron-electron interactions are mediated
via the quantum electrodynamic field, but the relevant wavelengths are
substantially shorter than the ones dominating the Casimir interaction
of different objects, so we can safely approximate the short
wavelength interactions by effective potentials $u(\vecx)$ and
$v(\vecx-\vecx')$.

To compute the total partition function, we exponentiate the time
integral of the Lagrangian and analytically continue
the time coordinate $t$ to $-i\tau$, yielding
\be
Z(\beta) = \int \dA \, e^{-\frac{1}{\hbar}S_\text{EM}}
Z_\text{elec}[\A],
\label{eq:Ztot}
\ee
where
\be
S_\text{EM} = \thalf \int_0^{\hbar\beta} d\tau \int d\vecx \,
\left(\E^2+\B^2\right),
\ee
is the free electromagnetic action and
\be
Z_\text{elec}[\A] = \int \mathcal{D}\bar{\psi}\mathcal{D}\psi
\, e^{- \frac{1}{\hbar}S_\text{elec}}
\ee
is the electronic partition function.  The Euclidean electronic action is
\be
S_\text{elec} = \int_0^{\hbar \beta} d\tau \int d\vecx \,
\left(
\hbar \bar{\psi}\partial_\tau \psi + \frac{\hbar^2}{2m}
\left[\left(\nabla-\frac{i e}{\hbar} \A\right)\bar{\psi}\right]
\left[\left(\nabla+\frac{i e}{\hbar} \A\right)\psi\right]
+ \hbox{interactions}\right),
\ee
where the interaction terms are taken from 
\refeqn{elecinter}, including at least  
$\hat{\mathcal{L}}_\text{lattice}$ to keep the electrons confined to
the various objects.  Here the creation and annihilation operators
$\hat{\psi}^\dagger$ and $\hat{\psi}$ go over to Grassman path integral
variables $\bar{\psi}$ and $\psi$.

Next, we expand the partition function of the electronic system,
\be
Z_\text{elec}[\A] \approx Z_\text{elec}[0] 
\left(1 + \thalf \sum_{i,j} \int_0^{\hbar \beta} d\tau d\tau' 
\int d\vecx d\vecx' K_{ij}(\tau-\tau',\vecx,\vecx')
A^i(\tau,\vecx) A^j(\tau',\vecx')\right),
\label{eq:Zelec}
\ee
where the second derivative of $Z_\text{elec}$,
\be
K_{ij}(\tau-\tau', \vecx,\vecx') =
\left.\frac{1}{Z_\text{elec}[0]}\frac{\delta^2
Z_\text{elec}[\A]}{\delta A^i(\tau,\vecx) \delta
A^j(\tau',\vecx')}\right|_{\A=0},
\label{eq:K}
\ee
only depends on the difference in imaginary time $\tau-\tau'$. The
linear term has been omitted in this expansion because it vanishes
for systems with no mean currents.  We then obtain the well-known Kubo
formula for electrical conductivity, \cite{Sachdev99}
\be
\sigma^T_{ij}(ic\kappa_n,\vecx,\vecx') = - \frac{1}{c\kappa_n}
\int_0^{\hbar \beta} d\tau \,  K_{ij}(\tau, \vecx,\vecx') 
e^{ic\kappa_n \tau}.
\label{eq:sigma}
\ee
The $T$ superscript indicates that this is the imaginary-time ordered
response function. The retarded response function can be obtained by
the substitution $ic\kappa_n \to \omega + i 0^+$. The conductivity is
related to the permittivity by
\be
\sigma^T_{ij}(ic\kappa_n,\vecx,\vecx') = 
c\kappa_n\left(\epsilon_{ij}^T(ic\kappa_n,\vecx,\vecx') - 
\delta_{ij}\delta(\vecx-\vecx')\right).
\ee

After substituting into \refeqn{Zelec}, we obtain
\be
\begin{split}
Z_\text{elec} \approx Z_\text{elec}[0]
\left(1 + \frac{\beta}{2} \sum_{n=-\infty}^\infty \int d\vecx d\vecx'
\E^*(ic\kappa_n,\vecx)\cdot
\left(\mathbb{I}\delta(\vecx-\vecx') -
\boldsymbol{\epsilon}^T (ic\kappa_n,\vecx,\vecx')\right)
\E(ic\kappa_n,\vecx')\right) 
\end{split}
\ee
and, finally, after reexponentiating and plugging into
\refeqn{Ztot} we obtain
\be
\begin{split}
Z(\beta) \approx Z_\text{elec}[0] \int \dA
\exp & \left[-\beta\sum_{n=-\infty}^\infty \left(
\int d\vecx d\vecx' \,
\E^*(ic\kappa_n,\vecx)\cdot
\boldsymbol{\epsilon}^T(ic\kappa_n,\vecx,\vecx')\E(ic\kappa_n,\vecx')
\right. \right. \\ & \left. \left. +
\int d\vecx \, \B^*(ic\kappa_n,\vecx)\cdot \B(ic\kappa_n,\vecx)
\right)\right].
\end{split}
\ee

The imaginary-time ordered response function can be obtained from the
retarded real time response function, which is experimentally
accessible, by analytic continuation to imaginary frequencies. The
relationship $\epsilon^T_{ij}(ic\kappa_n,\vecx,\vecx') =
\epsilon_{ij}^R(ic|\kappa_n|,\vecx,\vecx')$ between the two response
functions and the symmetry of the retarded response function in the
indices $(i,\vecx)$ and $(j,\vecx')$ is discussed in
Refs. \cite{LandauLifshitzS180,Abrikosov75}.

To be able to consider two objects as separate and distinct, we
assume that $\epsilon^T(ic\kappa_n,\vecx,\vecx')$ is zero when
$\vecx$ and $\vecx'$ are on different objects, separated by the
vacuum or a medium. This is justified even for small gaps between the
objects, since tunneling probabilities decay exponentially in the gap
distance.

Although the dissipative part of the electric response (the
conductivity) appears in the above equations, there is not actual
dissipation at zero temperature.  Rather,
Eqs. (\ref{eq:Zelec})--(\ref{eq:sigma}) merely show that the
conductivity $\sigma^R$, which can be measured in an experiment, can
be related to the fluctuations that exist in a system in the absence
of an external perturbation. The size of these fluctuations is
represented by the second derivative of the partition function,
\refeqn{K}.

Although the magnetic susceptibility is usually negligible compared
to the electric susceptibility, we have kept the permeability function
$\mu$ in our derivations of Casimir interactions. If the spin-magnetic
field coupling term $\hat{\mathcal{L}}_\text{spin}$ replaces or
dominates the standard coupling between charge and electromagnetic
field in \refeqn{L}, an analogous procedure introduces
the magnetic permeability function into the action. In that case, the
partition function of the matter field, the analogue of
\refeqn{Zelec}, has to be expanded in
$\B^i=(\boldsymbol{\nabla}\times\A)^i$ instead of $\A^i$.

\section{Green's function expansions and modified eigenfunctions}
\label{sec:Greenexp}
In this section we supply the Green's function expansions
for imaginary frequency $ic\kappa$ and the associated
modified eigenfunctions in various bases.  After
analytically continuing the wave functions in \refeqn{G0expem}, it is
convenient to redefine them by multiplication by an overall factor
in order to obtain the conventional definitions of the modified wave
functions.  Since $\Eincc(\kappa,\vecx)= \Eout(\kappa,\vecx)$, it
suffices to supply the modified regular and outgoing wave
functions.

Electromagnetic vector waves are typically divided into TE and
TM modes. It is customary to name the TE wave functions
$\M(\omega,\vecx)$ and the TM waves $\N(\omega,\vecx)$. 
Often TE modes are referred to as magnetic modes, labeled by $M$, and
TM modes are referred to as electric modes, labeled by $E$. 

\subsection{Green's function and eigenfunctions -- plane wave basis}

We choose the $\hatz$ axis as a convenient symmetry axis and let
$\veckpe$ denote momentum perpendicular to this axis.  The free dyadic
Green's function is
\be
\begin{split}
& \tGzero(ic\kappa,\vecx,\vecx') = \\
& \hspace{-6pt}
\int \frac{d\veckpe}{(2\pi)^2}
\left\{
\begin{array}{l l}
C_{\veckpe M}(\kappa)
\Mout_{\veckpe}(\kappa,\vecx) \otimes \Mrcc_{\veckpe}(\kappa,\vecx')
+
C_{\veckpe E}(\kappa)
\Nout_{\veckpe}(\kappa,\vecx) \otimes \Nrcc_{\veckpe}(\kappa,\vecx')
& \text{if } z > z' \\
C_{\veckpe M}(\kappa)
\Mr_{\veckpe}(\kappa,\vecx) \otimes \Mincc_{\veckpe}(\kappa,\vecx')
+
C_{\veckpe E}(\kappa)
\Nr_{\veckpe}(\kappa,\vecx) \otimes \Nincc_{\veckpe}(\kappa,\vecx')
& \text{if } z < z'
\end{array}
\right..
\end{split}
\labeleqn{tG0plane}
\ee
Here, $C_{\veckpe M}(\kappa) = \frac{1}{2\sqrt{\veckpe^2+\kappa^2}} = (-1)
C_{\veckpe E}(\kappa)$, and the modified vector plane wave functions
are given by
\be
\begin{split}
\Mr_{\veckpe}(\kappa,\vecx) & = \frac{1}{|\veckpe|} \curl
\phir_{\veckpe}(\kappa,\vecx) \hatz, \qquad
\Mout_{\veckpe}(\kappa,\vecx) = \frac{1}{|\veckpe|} \curl
\phiout_{\veckpe}(\kappa,\vecx)
\hatz, \\
\Nr_{\veckpe}(\kappa,\vecx)   & = \frac{1}{\kappa|\veckpe|} \curl
\curl \phir_{\veckpe}(\kappa,\vecx)
\hatz, \qquad
\Nout_{\veckpe}(\kappa,\vecx) = \frac{1}{\kappa|\veckpe|} \curl \curl
\phiout_{\veckpe}(\kappa,\vecx) \hatz ,
\end{split}
\labeleqn{G0emplanedetails}
\ee
in terms of the modified scalar plane wave functions,
\be
\phir_{\veckpe}(\kappa,\vecx) = e^{i \veckpe\cdot\vecxpe +
\sqrt{\veckpe^2+\kappa^2} z }, \qquad
\phiout_{\veckpe}(\kappa,\vecx) = e^{i \veckpe\cdot\vecxpe-
\sqrt{\veckpe^2+\kappa^2} z}.
\ee
As discussed in the text, the labels ``reg,'' ``out,'' and ``in'' are
not really appropriate for plane wave functions, but the mathematical
results all carry over.

\subsection{Green's function and eigenfunctions -- cylindrical wave basis}

Again, let us take the $\hatz$ axis as a convenient symmetry axis and
let $\rho = \sqrt{x^2+y^2}$ be the distance to the $\hatz$ axis.  The
free dyadic Green's function is given by
\be
\begin{split}
& \tGzero(ic\kappa,\vecx,\vecx') = \\
& 
\int \frac{dk_z}{2\pi} \sum_n
\left\{
\begin{array}{l l}
C_{M}
\Mout_{k_z n}(\kappa,\vecx) \otimes \Mrcc_{k_z n}(\kappa,\vecx')
+
C_{E}
\Nout_{k_z n}(\kappa,\vecx) \otimes \Nrcc_{k_z n}(\kappa,\vecx')
& \text{if } \rho > \rho' \\
C_{M}
\Mr_{k_z n}(\kappa,\vecx) \otimes \Mincc_{k_z n}(\kappa,\vecx')
+
C_{E}
\Nr_{k_z n}(\kappa,\vecx) \otimes \Nincc_{k_z n}(\kappa,\vecx')
& \text{if } \rho < \rho'
\end{array}
\right. .
\end{split}
\labeleqn{tG0cyl}
\ee
Here $C_{E}=\frac{1}{2\pi} = (-1)C_{M}$,
the vector cylindrical wave functions are given by
\be
\begin{split}
\Mr_{k_z n}(\kappa,\vecx)   & = \frac{1}{\sqrt{k_z^2+\kappa^2}}
\curl \phir_{k_z n}(\kappa,\vecx) \hatz, \\
\Mout_{k_z n}(\kappa,\vecx) & = \frac{1}{\sqrt{k_z^2+\kappa^2}}
\curl \phiout_{k_z n}(\kappa,\vecx) \hatz, \\
\Nr_{k_z n}(\kappa,\vecx)   & = \frac{1}{\kappa \sqrt{k_z^2+\kappa^2}}
\curl \curl \phir_{k_z n}(\kappa,\vecx) \hatz, \\
\Nout_{k_z n}(\kappa,\vecx) & = \frac{1}{\kappa \sqrt{k_z^2+\kappa^2}}
\curl \curl \phiout_{k_z n}(\kappa,\vecx) \hatz,
\end{split}
\ee
and the modified cylindrical wave functions are
\be
\begin{split}
\phir_{k_z n}(\kappa, \vecx) = I_n\left(\rho\sqrt{k_z^2+\kappa^2}\right)
e^{ik_z z+in\theta}, \qquad
\phiout_{k_z n}(\kappa, \vecx) = K_n\left(\rho\sqrt{k_z^2+\kappa^2}\right)
e^{ik_z z+in\theta},
\end{split}
\ee
where $I_n$ is the modified Bessel function of the first kind and
$K_n$ is the modified Bessel function of the third kind.

\subsection{Green's function and eigenfunctions -- spherical wave basis}

In spherical coordinates the free dyadic Green's function is given by
\be
\begin{split}
& \tGzero(ic\kappa,\vecx,\vecx') = \\
& 
\sum_{lm}
\left\{
\begin{array}{l l}
C_{M}(\kappa)
\Mout_{l m }(\kappa,\vecx) \otimes \Mrcc_{l m }(\kappa,\vecx')
+
C_{E}(\kappa)
\Nout_{l m }(\kappa,\vecx) \otimes \Nrcc_{l m }(\kappa,\vecx')
& \text{if } |\vecx| > |\vecx'| \\
C_{M}(\kappa)
\Mr_{l m }(\kappa,\vecx) \otimes \Mincc_{l m }(\kappa,\vecx')
+
C_{E}(\kappa)
\Nr_{l m }(\kappa,\vecx) \otimes \Nincc_{l m }(\kappa,\vecx')
& \text{if } |\vecx| < |\vecx'|
\end{array}
\right. .
\end{split}
\labeleqn{tG0sphere}
\ee
Here $C_{M}(\kappa) = \kappa = (-1) C_{E}(\kappa)$, 
the vector spherical wave functions are
\be
\begin{split}
\Mr_{lm}(\kappa,\vecx)   & = \tfrac{1}{\sqrt{l(l+1)}}
\curl \phir_{lm}(\kappa,\vecx) \vecx, \qquad
\Mout_{lm}(\kappa,\vecx) = \tfrac{1}{\sqrt{l(l+1)}}
\curl \phiout_{lm}(\kappa,\vecx) \vecx, \\
\Nr_{lm}(\kappa,\vecx)   & = \tfrac{1}{\kappa \sqrt{l(l+1)}}
\curl \curl \phir_{lm}(\kappa,\vecx) \vecx, \qquad
\Nout_{lm}(\kappa,\vecx) = \tfrac{1}{\kappa \sqrt{l(l+1)}}
\curl \curl \phiout_{lm}(\kappa,\vecx) \vecx,
\end{split}
\labeleqn{G0emspheredetails}
\ee
and the modified spherical wave functions are
\be
\begin{split}
\phir_{lm}(\kappa,\vecx)  = i_l(\kappa |\vecx|) Y_{lm}(\hat{\vecx}), \qquad
\phiout_{lm}(\kappa,\vecx) = k_l(\kappa |\vecx|) Y_{lm}(\hat{\vecx}),
\end{split}
\ee
where $i_l(z)=\sqrt{\frac{\pi}{2 z}} I_{l+1/2}(z)$ is the modified
spherical Bessel function of the first kind, and
$k_l(z)=\sqrt{\frac{2}{\pi z}} K_{l+1/2}(z)$ is the modified spherical
Bessel function of the third kind.

\subsection{Green's function -- elliptic cylindrical basis}
\label{ellipticcylinder}

In order to study geometry and orientation dependence of Casimir
interactions, it is helpful to be able to study objects
with reduced symmetry.  In Ref. \cite{Emig09}, this formalism was applied
to spheroids in scalar field theory.  Unfortunately, the vector
Helmholtz equation is not separable in spheroidal coordinates as it is
in spherical coordinates.  While the analogous vector spheroidal
harmonics can still be constructed, the scattering matrix for a
perfectly conducting spheroid is not diagonal, although it can be
obtained from a more elaborate calculation \cite{Schulz}.  For a
perfectly conducting elliptic cylinder, however, the vector scattering
problem is separable, so we describe that case here.  Throughout this
section, we use the same normalization and conventions as in Ref. 
\cite{Graham:2005cq}, in which all functions in elliptic cylindrical
coordinates have the same normalization as their circular analogs.  As
a result, all the functions inherit the usual completeness and
orthonormality relations and approach their circular analogs in the
limit of long wavelength.

In elliptic cylindrical coordinates, the $z$ coordinate is
unchanged, while the components of $\vecxpe$ become
$x=a\cosh\mu\cos\theta$ and $y = a\sinh\mu\sin\theta$,
where the interfocal separation of the ellipse is $2a$.  Far away,
$\theta$ approaches the ordinary angle in cylindrical coordinates and
$|\vecxpe| \approx \frac{a}{2} e^\mu$.  Separation of variables in these
coordinates yields angular and radial Mathieu functions for $\theta$
and $\mu$, respectively.  The even and odd angular Mathieu functions
are  $ce_n(\theta,\gamma)$ with $n\geq 0$ and $se_n(\theta,\gamma)$
with $n > 0$, which are the analogs of $\cos n\theta$ and $\sin n
\theta$ in the circular case.  (We used a complex exponential basis
for the circular case, but it could equally well be represented in
terms of sines and cosines.)  The angular functions now depend on the
wave number through the combination $\gamma=-(k_z^2 +\kappa^2)
a^2/2$.  The corresponding radial functions are now different for the
even and odd cases and depend on $\gamma$ and $\mu$ separately rather than
through a single product of the two.  The even and odd modified  
radial Mathieu functions of the first kind are denoted 
$Ie_m(\mu, \gamma)$ and $Io_m(\mu, \gamma)$ respectively, and 
the even and odd modified radial Mathieu functions of the third kind
are denoted  $Ke_m(\mu, \gamma)$ and $Ko_m(\mu, \gamma)$, respectively.

We then obtain the same results as in cylindrical coordinates, but now
with 
\be
\begin{split}
\phir_{k_z n e}(\kappa,\vecx) = 
Ie_n(\mu, \gamma) ce_n(\theta, \gamma) e^{ik_z z}, \qquad
\phiout_{k_z n e}(\kappa,\vecx) &= 
Ke_n(\mu, \gamma) ce_n(\theta, \gamma) e^{ik_z z}, \\
\phir_{k_z n o}(\kappa,\vecx) = 
Io_n(\mu, \gamma) se_n(\theta, \gamma) e^{ik_z z}, \qquad
\phiout_{k_z n o}(\kappa,\vecx) &= 
Ko_n(\mu, \gamma) se_n(\theta, \gamma) e^{ik_z z}.
\end{split}
\ee
For numerical calculation the required Mathieu functions can be
efficiently computed using the C++ package of Alhargan
\cite{Alhargan:2000,Alhargan:2000a}.
Analogous replacements convert the translation matrices
and wave conversion matrices described below into this basis.

\section{Translation matrices}
\label{sec:Translation}

In the following, we list the translation matrices that make up 
$\X^{ij}$, defined in \refeqn{Xdef}. The definition of the vector
$\vecX_{ij}$, which points from the origin of object $i$ to the origin
of object $j$, is illustrated in \reffig{transboth}.

\subsection{Plane wave basis}

Plane waves are eigenfunctions of the translation operator,
which does not mix TE and TM vector plane wave functions.

If the $z$ coordinates of object $i$ are smaller than those of object
$j$, then $-\mathcal{V}^{ij}$ is the only nonzero entry in $\X^{ij}$.  
Taking $\vecX_{ij}$ to point from the origin of object $i$, $\orig_i$, to
the origin of object $j$, $\orig_j$ (that is, upward), we obtain
\be
\begin{split}
\mathcal{V}^{ij}_{\veckpe P, \veckpe' P'}  & = 
e^{- i\veckpe \cdot \vecX_{ij,\perp} - \sqrt{\veckpe^2+\kappa^2}X_{ij,z} }
\tfrac{(2\pi)^2}{L^2}\delta^{(2)}(\veckpe-\veckpe')\delta_{P,P'}.
\end{split}
\ee

If $i$ is located above $j$, then $-\mathcal{W}^{ji}$ is the only
nonzero entry in $\X^{ij}$. The vector $\vecX_{ji}$ points upward from
$\orig_j$ to $\orig_i$, and we have
\be
\begin{split}
\mathcal{W}^{ji}_{\veckpe P,\veckpe' P'} & = 
\mathcal{V}^{ji*}_{\veckpe' P',\veckpe P} 
\frac{C_{\veckpe P}(\kappa)}{C_{\veckpe' P'}(\kappa)} \\
& = e^{i\veckpe \cdot \vecX_{ji,\perp} -
\sqrt{\veckpe^2+\kappa^2} X_{ji,z}}
\tfrac{(2\pi)^2}{L^2}\delta^{(2)}(\veckpe-\veckpe') \delta_{P,P'}.
\end{split}
\ee
Since the matrix is diagonal in $\veckpe$ and $P$, the factor
$\frac{C_{\veckpe P}(\kappa)}{C_{\veckpe' P'}(\kappa)}$ cancels.

\subsection{Cylindrical wave basis}

Translations do not mix the TE and TM modes of vector cylindrical wave
functions. They are constructed by taking the scalar cylindrical wave
function, multiplying by the unit vector $\hatz$, and performing one or
two curl operations.  A TE vector cylindrical wave function is
perpendicular to $\hatz$, while the curl of a TM vector cylindrical
wave function is perpendicular to $\hatz$. Expanding any of the two
vector wave functions around any other point in space must preserve
its orthogonality property with respect to the constant vector
$\hatz$. So, the two are not mixed by the translation matrix.

If two objects $i$ and $j$ are outside of one another,
$-\mathcal{U}^{ji}$ is the only nonzero submatrix of $\X^{ij}$. Again, let
$\vecX_{ji}$ point from $\orig_j$ to $\orig_i$.  We have
\be
\begin{split}
\mathcal{U}^{ji}_{k_z n P, k_z' n' P'} 
& = K_{n-n'}\left(|\vecX_{ji,\perp}|\sqrt{k_z^2+\kappa^2}\right)
e^{-i k_z X_{ji,z}-i(n-n')\theta_{ji}}(-1)^{n'} \delta_{P,P'}
\tfrac{2\pi}{L}\delta(k_z-k_z'),
\end{split}
\ee
where $|\vecX_{ji,\perp}|$ is the distance of $\vecX_{ji}$ to the $\hatz$
axis, {\it i.e.\/} the length of the projection onto the $x$-$y$ plane, and
$\theta_{ji}$ is the angle of $\vecX_{ji}$ in the $x$-$y$ plane.

When object $i$ is enclosed inside the surface of an infinite
cylinder, inside object $j$, submatrix $-\mathcal{V}^{ij}$ is the only
nonzero entry in $\X^{ij}$.  We have
\be
\begin{split}
\mathcal{V}^{ij}_{k_z n P, k_z' n' P'}
& = I_{n-n'}\left(|\vecX_{ij,\perp}|\sqrt{k_z^2+\kappa^2}\right)
e^{-i k_z X_{ij,z}-i(n-n')\theta_{ij}}(-1)^{n+n'} \delta_{P,P'}
\tfrac{2\pi}{L}\delta(k_z-k_z'),
\end{split}
\ee
where $\vecX_{ij}$ points from $\orig_i$ to $\orig_j$,

If the roles of $i$ and $j$ are reversed, then $-\mathcal{W}^{ji}$ is the
nonzero submatrix of $\X^{ij}$, with
\be
\begin{split}
\mathcal{W}^{ji}_{k_z n P, k_z' n' P'} & = 
\mathcal{V}^{ji*}_{k_z' n' P', k_z n P} \frac{C_{P}}{C_{P'}} \\
& = I_{n-n'}\left(|\vecX_{ji,\perp}|\sqrt{k_z^2+\kappa^2}\right)
e^{+i k_z X_{ji,z}-i(n-n')\theta_{ji}}(-1)^{n+n'} \delta_{P,P'}
\tfrac{2\pi}{L}\delta(k_z-k_z').
\end{split}
\ee
Since the matrix is diagonal in $P$, the factor $\frac{C_{P}}{C_{P'}}$
cancels.

\subsection{Spherical wave basis}

The TE vector wave functions are orthogonal to the radius vector
$\vecx$.  Since the same vector wave function cannot also be orthogonal
everywhere to the radius vector of a shifted coordinate system, TE
and TM polarizations mix under translation.

Suppose object $i$ and its origin are outside a spherical separating
surface, which encloses $j$. The nonzero submatrix of $\X^{ij}$ is
$-\mathcal{U}^{ji}$, with
\be
\begin{split}
\mathcal{U}^{ji}_{l' m'  M, l m M}
& = 
(-1)^{m+l}  \sum_{l''}
\left[l(l+1)+l'(l'+1)-l''(l''+1)\right]
    \sqrt{\frac{\pi(2l+1)(2l'+1)(2l''+1)}{l(l+1)l'(l'+1)}} \\
&\times\begin{pmatrix}l&l'&l''\\0&0&0\end{pmatrix}
\begin{pmatrix}l&l'&l''\\m&-m'&m'-m\end{pmatrix}
k_{l''}(\kappa|\vecX_{ji}|) Y_{l''m-m'}(\hat \vecX_{ji}), \\
\mathcal{U}^{ji}_{l' m' E, l m M}
& = 
-\frac{i\kappa}{\sqrt{l(l+1)l'(l'+1)}} \, \vecX_{ji} \cdot \bigg[
\hat \vecx \frac{1}{2} \left(\lambda^+_{lm} A_{l'm'lm+1}(\vecX_{ji}) +
\lambda^-_{lm} A_{l'm'lm-1}(\vecX_{ji}) \right)\nonumber\\
&+ \, \hat \vecy \frac{1}{2i} \left(\lambda^+_{lm} 
A_{l'm'lm+1}(\vecX_{ji}) -
\lambda^-_{lm} A_{l'm'lm-1}(\vecX_{ji}) \right)
+ \hat \vecz\, m \,A_{l'm'lm}(\vecX_{ji})
\bigg] \, ,\\
\mathcal{U}^{ji}_{l' m' M, l m E} & = - \mathcal{U}^{ji}_{l' m' E, l m M}, 
\qquad
\mathcal{U}^{ji}_{l' m' E, l m E} = \mathcal{U}^{ji}_{l' m' M, l m M},
\end{split}
\ee
where
\begin{eqnarray}
  \label{eq:A-mtrix-elements}
A_{l'm'lm}(\vecX_{ji}) &=& (-1)^{m+l}  \sum_{l''}
\sqrt{4\pi(2l+1)(2l'+1)(2l''+1)} \nonumber\\
&&\times \begin{pmatrix}l&l'&l''\\0&0&0\end{pmatrix}
\begin{pmatrix}l&l'&l''\\m&-m'&m'-m\end{pmatrix}
k_{l''}(\kappa|\vecX_{ji}|) Y_{l''m-m'}(\hat \vecX_{ji})  
\end{eqnarray}
and $\lambda^\pm_{lm}=\sqrt{(l\mp m)(l\pm m+1)}$.

The translations between regular waves are described by the matrix elements
\be
\begin{split}
\mathcal{V}^{ij}_{l' m' M, l m M} 
& = 
(-1)^{m}  \sum_{l''}
\left[l(l+1)+l'(l'+1)-l''(l''+1)\right]
    \sqrt{\frac{\pi(2l+1)(2l'+1)(2l''+1)}{l(l+1)l'(l'+1)}} \\
&\times\begin{pmatrix}l&l'&l''\\0&0&0\end{pmatrix}
\begin{pmatrix}l&l'&l''\\m&-m'&m'-m\end{pmatrix}
i_{l''}(\kappa|\vecX_{ij}|) (-1)^{l''} Y_{l''m-m'}(\hat \vecX_{ij}), \\
\mathcal{V}^{ij}_{l' m' E, l m M} 
& = 
-\frac{i\kappa}{\sqrt{l(l+1)l'(l'+1)}} \, \vecX_{ij} \cdot \bigg[
\hat \vecx \frac{1}{2} \left(\lambda^+_{lm} B_{l'm'lm+1}(\vecX_{ij}) +
\lambda^-_{lm} B_{l'm'lm-1}(\vecX_{ij}) \right)\nonumber\\
&+ \, \hat \vecy \frac{1}{2i} \left(\lambda^+_{lm} 
B_{l'm'lm+1}(\vecX_{ij}) -
\lambda^-_{lm} B_{l'm'lm-1}(\vecX_{ij}) \right)
+ \hat \vecz\, m \,B_{l'm'lm}(\vecX_{ij})
\bigg] \, ,\\
\mathcal{V}^{ij}_{l' m' M, l m E} & = - \mathcal{V}^{ij}_{l' m' E, l m M},
\qquad
\mathcal{V}^{ij}_{l' m' E, l m E} = \mathcal{V}^{ij}_{l' m' M, l m M},
\end{split}
\ee
where
\begin{eqnarray}
  \label{eq:B-mtrix-elements}
B_{l'm'lm}(\vecX_{ij}) &=& (-1)^{m}  \sum_{l''}
\sqrt{4\pi(2l+1)(2l'+1)(2l''+1)} \nonumber\\
&&\times \begin{pmatrix}l&l'&l''\\0&0&0\end{pmatrix}
\begin{pmatrix}l&l'&l''\\m&-m'&m'-m\end{pmatrix}
i_{l''}(\kappa|\vecX_{ij}|) (-1)^{l''} Y_{l''m-m'}(\hat \vecX_{ij})  
\end{eqnarray}
and $\lambda^\pm_{lm}=\sqrt{(l\mp m)(l\pm m+1)}$.

The matrix $\mathcal{W}^{ji}$ is related to $\mathcal{V}^{ji}$,
\be
\mathcal{W}^{ji}_{l' m' P', l m P} =
\mathcal{V}^{ji\dagger}_{l' m' P', l m P}
\frac{C_{P'}(\kappa)}{C_{P}(\kappa)}.
\ee

$\mathcal{V}^{ji}$, of course, is the same as $\mathcal{V}^{ij}$ 
with $\vecX_{ij}$
replaced by $\vecX_{ji}$. To be more specific, the elements correspond
in the following way,
\be
\begin{split}
\mathcal{W}^{ji}_{l'm'M,lmM} = \mathcal{V}^{ji *}_{lmM,l'm'M}\,, & \quad
\mathcal{W}^{ji}_{l'm'E,lmM} = -\mathcal{V}^{ji *}_{lmM,l'm'E}\,, \\
\mathcal{W}^{ji}_{l'm'M,lmE} = -\mathcal{V}^{ji *}_{lmE,l'm'M}\,, & \quad
\mathcal{W}^{ji}_{l'm'E,lmE} = \mathcal{V}^{ji *}_{lmE,l'm'E}\,.
\end{split}
\ee

\section{Wave conversion matrices}
\label{app:Conversion}
It is not necessary to express all the objects' scattering amplitudes
in the same basis. Here, we supply the matrices that convert modified
vector plane wave functions to spherical or cylindrical vector wave functions.

\subsection{Vector plane wave functions to spherical vector wave functions}

\be
\begin{split}
\Mr_{\veckpe}(\kappa,\vecx) & = \sum_{lm} D_{lmM,\veckpe M}
\Mr_{lm}(\kappa,\vecx) + D_{lmE,\veckpe M} \Nr_{lm}(\kappa,\vecx)\, , \\
\Nr_{\veckpe}(\kappa,\vecx) & = \sum_{lm} D_{lmM,\veckpe E}
\Mr_{lm}(\kappa,\vecx) + D_{lmE,\veckpe E} \Nr_{lm}(\kappa,\vecx).
\end{split}
\labeleqn{convMNsphereplane}
\ee
The conversion matrices are obtained from the decomposition of a plane
wave in spherical coordinates,
\be
\begin{split}
D_{lmM,\veckpe M} & = \sqrt{\frac{4\pi(2l+1)(l-m)!}{l(l+1)(l+m)!}}
\frac{|\veckpe|}{\kappa} e^{-im\phi_{\veckpe}} P_{l}^{\prime m}
\left(\sqrt{\veckpe^2+\kappa^2}/\kappa\right) \, ,\\
D_{lmE,\veckpe M} & = \sqrt{\frac{4\pi(2l+1)(l-m)!}{l(l+1)(l+m)!}}
i m \frac{\kappa}{|\veckpe|} e^{-im\phi_{\veckpe}} P_{l}^{m}
\left(\sqrt{\veckpe^2+\kappa^2}/\kappa\right) \, , \\
D_{lmE,\veckpe E} & = D_{lm M,\veckpe M}, \qquad
D_{lm M,\veckpe E} = -D_{lm E,\veckpe M},
\end{split}
\labeleqn{convDsphereplane}
\ee
where $P_l^{m}$ is the associated Legendre polynomial and prime
indicates the derivative of $P_l^m$ with respect to its argument.

\subsection{Vector plane wave functions to cylindrical vector wave functions}

The cylindrical vector wave functions are defined as before, but 
now we consider regular vector plane wave functions
that decay along the $-\mathbf{\hat{x}}$ axis instead of the
$-\hatz$ axis, 
\be
\begin{split}
\Mr_{\veckpe}(\kappa,\vecx) & = \frac{1}{\sqrt{k_y^2+k_z^2}} \curl
e^{\sqrt{\kappa^2+k_y^2+k_z^2} x + i k_y y + i k_z z} \mathbf{\hat{x}}, \\
\Nr_{\veckpe}(\kappa,\vecx)   & = \frac{1}{\kappa\sqrt{k_y^2+k_z^2}} \curl
\curl e^{\sqrt{\kappa^2+k_y^2+k_z^2} x + i k_y y + i k_z z} \mathbf{\hat{x}}.
\end{split}
\ee

The vector plane wave functions can be decomposed in vector
cylindrical wave functions,
\be
\begin{split}
\Mr_{\veckpe}(\kappa,\vecx) & = \sum_{n} D_{k_z n M,\veckpe M}
\Mr_{k_z n}(\kappa,\vecx) + D_{k_z n E,\veckpe M} 
\Nr_{k_z n}(\kappa,\vecx) \, , \\
\Nr_{\veckpe}(\kappa,\vecx) & = \sum_{n} D_{k_z n M,\veckpe E}
\Mr_{k_z n}(\kappa,\vecx) + D_{k_z n E,\veckpe E} 
\Nr_{k_z n}(\kappa,\vecx) \, ,
\end{split}
\labeleqn{convMNcylinderplane}
\ee
using the conversion matrix elements
\be
\begin{split}
D_{k_z n M, \veckpe M} & = -i \frac{k_z}{\sqrt{k_y^2+k_z^2}}
\sqrt{1+\xi^2} \left(\sqrt{1+\xi^2}+\xi\right)^n \, ,\\
D_{k_z n E, \veckpe M} & = i \frac{\kappa}{\sqrt{k_y^2+k_z^2}}
\xi \left(\sqrt{1+\xi^2}+\xi\right)^n \, , \\
D_{k_z n E, \veckpe E} & = D_{k_z n M, \veckpe M} \, , \qquad
D_{k_z n M, \veckpe E} = -D_{k_z n E, \veckpe M},
\end{split}
\ee
where $\xi = \frac{k_y}{\sqrt{\kappa^2+k_z^2}}$ and $\veckpe=(k_y,k_z)$.

\end{appendix}

\end{document}